\documentclass[prb,twocolumn,showpacs,aps]{revtex4}
\usepackage{graphicx}

\def\imat  { i}

\begin{document}

\title{Squeezing of a nanomechanical resonator
by quantum \\
nondemolition measurement and feedback}

% QND squeezing of a nanomechanical resonator via continuous stroboscopic
%   measurement and feedback
% Squeezing nanoresonator by quantum nondemolition measurements and feedback

\author{Rusko Ruskov,$^1$\footnote{On leave of absence from Institute
for Nuclear Research and Nuclear Energy, Sofia BG-1784, Bulgaria;
Present address:
Physics Department, Pennsylvania State University, University Park, PA 16802}
Keith Schwab,$^2$ and Alexander N. Korotkov$^1$}
\affiliation{$^1$Department of Electrical Engineering, University of
California, Riverside, CA 92521 \\
$^2$Laboratory for Physical Sciences, 8050 Greenmead Drive,
College Park, MD 20740 }

\date{\today}

\begin{abstract}
      We analyze  squeezing of the nanoresonator state produced
by periodic measurement of position by
a quantum point contact or a single-electron transistor.
The mechanism of squeezing is the stroboscopic quantum nondemolition
measurement generalized to the case of continuous
measurement by a weakly coupled detector. The magnitude of squeezing
is calculated for the harmonic and stroboscopic modulations of measurement,
taking into account detector efficiency and nanoresonator quality factor.
We also analyze the operation of the quantum feedback, which prevents
fluctuations of the wavepacket center due to measurement back-action.
Verification of the squeezed state can be performed in almost the same
way as its preparation; similar procedure can also be used for the force
detection with sensitivity beyond the standard quantum limit.
\end{abstract}


\maketitle

\section{Introduction}

        Recent advances in fabrication of high-frequency nanomechanical
resonators \cite{ClelandRoukes,Craighead,Roukes-1GHz,Cleland-2003,Schwab-QL}
(see also  Refs.\ \onlinecite{Cleland-book} and \onlinecite{Blencowe-rev})
make possible the direct observation of their quantum behavior in the
nearest future.
Resonator frequency $\omega_0/2\pi$ slightly over 1 GHz has been already
demonstrated.\cite{Roukes-1GHz}
For such a resonator the condition $T < \hbar\omega_0$ (we use $k_B=1$)
is satisfied at temperature $T$ below $\sim 50$ mK, which is within
routine experimental range.
Actually, even in the case $T \gg \hbar\omega_0$
the quantum behavior is in principle observable  \cite{BraginskyKhalili}
if $T\tau_m/Q \lesssim \hbar $, where  $Q$ is the resonator
quality factor and $\tau_m$ is the typical measurement time.
This condition can be satisfied even for a MHz-range resonator
with large $Q$-factor, if measured with a good sensitivity
which translates into small $\tau_m$.
There is a rapid
experimental progress in monitoring the oscillating position of a
nanoresonator using radio-frequency single-electron transistor (RF-SET)
\cite{Cleland-2003,Schwab-QL} or quantum point contact (QPC)
\cite{Cleland-QPC} (at present RF-SET seems to be much more
efficient).
In particular, the position measurement accuracy $\Delta x$ within
the factor 5.8 from the standard quantum limit (SQL) $\Delta x_0$ has been
demonstrated; \cite{Schwab-QL} here $\Delta x_0 = \sqrt{\hbar /2m\omega_0}$
is the width (standard deviation) of the ground state of the oscillator
with mass $m$. Anticipating future progress in measurement precision,
in this paper we discuss a way of performing measurement with accuracy
better than $\Delta x_0$.

        Such measurement implies squeezing of the nanoresonator state
and requires using some tricks to avoid the effect of quantum back-action
from the detector which normally leads to the SQL.\cite{BraginskyKhalili}
Actually, an instantaneous position measurement by a strongly coupled
detector can in principle be made with precision $\Delta x$ better than
$\Delta x_0$ (orthodox projection, for example, implies $\Delta x =0$);
however, the limitation be the SQL arises for consecutive measurements
and also for  measurement by a weakly coupled detector, which
is necessarily continuous. The well-known way to overcome the SQL limitation
is to use quantum nondemolition (QND) measurements.
\cite{BraginskyKhalili,Braginsky2,Thorne,CavesRevModPhys}
The general idea of a QND measurement is to avoid
measuring (or obtaining any information on) the magnitude conjugated
to the magnitude of interest, and therefore to avoid the corresponding
back-action. An important implementation of this idea is the ``stroboscopic''
measurement of an oscillator position. \cite{Braginsky2,Thorne}
Suppose the position $x_1$ is measured (instantaneously) with a finite
precision $\Delta x$, which necessarily disturbs the momentum according
to the Heisenberg
uncertainty principle $\Delta p \geq \hbar /2\Delta x$. Normally this
momentum change would affect the result of the next position measurement
$x_2$ and would limit the accuracy for the position difference $x_2-x_1$,
leading to the SQL for this magnitude. However, if the second measurement
is performed exactly one oscillation period after the first one, the
oscillator returns to its initial state, and therefore the momentum change
does not affect the accuracy of $x_2-x_1$ measurement. Such stroboscopic
measurement gives no information related to the momentum, and this is exactly
the reason why the effect of quantum back-action is avoided.
\cite{BraginskyKhalili,Braginsky2,Thorne,CavesRevModPhys}

        The QND measurements have been mainly discussed in relation
to detection of very weak classical forces, in particular gravitational
waves (see, e.g., Refs.\ \onlinecite{Bocko,Kimble,Braginsky-2003}).
Recently the idea of QND measurements has been also applied to solid-state
mesoscopic structures (see, e.g., Refs.\ \onlinecite{Averin,Bulaevskii}).
Among other
recent developments (total number of papers on QND measurements is about
half a thousand) let us mention the experiment on atomic spin-squeezing
using the QND
measurement and real-time quantum feedback. \cite{Geremia}
Squeezing of a nanomechanical resonator using the QND measurement by
QPC or SET
and quantum feedback has been proposed in our recent Proceedings
paper; \cite{RSK} the present paper is a more complete
analysis of this proposal.

     Measurement of the nanoresonator position by the SET or QPC
has already received a significant theoretical attention.
\cite{Blencowe-rev,Blencowe,MozyrskyMartin,ArmourBlencoweSchwab,Smirnov,%
Schwab,Knobel,Hopkins,BlanterNazarov,Clerk-pd}
In particular, it was shown that
the process of measurement transfers the energy from the
detector to the nanoresonator leading to its ``heating''.
\cite{Blencowe,MozyrskyMartin}
A possible way to prevent such heating
is using the quantum feedback control of the nanoresonator
\cite{DohertyJacobs,Hopkins}
(other ideas for cooling have been proposed in Refs.\
\onlinecite{ShnirmanMartin} and \onlinecite{Wilson}).

The general idea of quantum feedback is very similar to classical feedback
and is based on the continuous monitoring of the system state and its
continuous control towards a desired state. However, the nontrivial part is
accurate monitoring of evolution of the quantum state (wavefunction
in the ideal case), which requires explicit account of the detector
back-action.
The quantum feedback of mesoscopic solid-state systems is a relatively
new subject, \cite{Ruskov-fb} though in quantum optics the quantum feedback
has been proposed more than a decade ago \cite{Wiseman} and has been
already realized experimentally. \cite{Geremia} The quantum feedback
analyzed in Ref.\ \onlinecite{Hopkins} assumes continuous monitoring
of the nanoresonator state with constant ``strength'' of measurement
and  allows cooling of the nanoresonator practically down to
the ground state. However, it does not allow squeezing of the nanoresonator
state (below $\Delta x_0$), except in unrealistic case of a strong coupling
between the nanoresonator and detector.

        Besides curiosity, the interest to nanoresonator squeezing is
justified by its importance for the ultrasensitive force detection.
Nanoresonator squeezing by periodic modulation of the spring constant
at twice the flexural frequency
has been proposed and analyzed in Refs.\ \onlinecite{Blencowe-sq} and
\onlinecite{Blencowe-rev} (this proposal is to some extent a scaled down
version of the proposal \cite{Grishchuk} for gravitational-wave detection
and experiment on classical thermomechanical noise squeezing
\cite{RugarGrutter}). Nanoresonator squeezing by reservoir engineering
(by coupling to a qubit and illumination with two microwaves) has been
proposed in Ref.\ \onlinecite{Rabl}. We would like to notice
that to be useful for an ultrasensitive force detection, the preparation
of a squeezed state should in any case be complemented by the measurement
stage after the force has acted on the nanoresonator; the most natural
way for this measurement is using the RF-SET (or QPC) as a detector,
and such measurement of a squeezed state is not trivial (unless
detector is strongly coupled).

\begin{figure}
\centering
\includegraphics[width=2.8in]{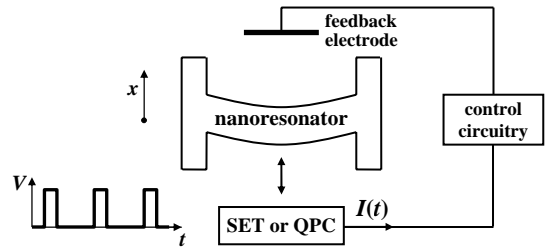}
% where an .eps filename suffix will be assumed under latex,
% and a .pdf suffix will be assumed for pdflatex
\caption{
Simplified schematic of the nanoresonator,
which position $x$ is measured by a single-electron transistor or a quantum
point contact. Stroboscopic modulation of the detector voltage $V(t)$
with frequency $2\,\omega_0/n$ leads to nanoresonator state squeezing.
Detector output $I(t)$ is used to monitor nanoresonator position. The
quantum feedback loop keeps the  center of the nanoresonator wavepacket
close to $x=0$.
}
\label{schematic}
\end{figure}

        In this paper we analyze the nanoresonator squeezing produced
by measuring the nanoresonator position (Fig.\ \ref{schematic})
with the measurement strength
modulated in time \cite{RSK} (for example, modulating the bias voltage
of the QPC or RF-SET), so the stages of the squeezed state preparation
and its measurement are essentially similar.
We show that even for a weak coupling with detector, a significant
squeezing of the nanoresonator state
can be achieved when the modulation frequency $\omega$ is close to
$2\,\omega_0/n$, $n=1, 2, \ldots$.

        The mechanism of this effect is exactly the physics of
stroboscopic QND measurements \cite{Braginsky2,Thorne} and can be
easily understood for the case of short measurement pulses applied
periodically, for example, once per oscillation period ($n=2$).
Each measurement pulse gives us some (though quite imprecise) information on
the nanoresonator position $x$ and correspondingly reduces the width of
the resonator density matrix in $x$-domain.
Between the pulses the resonator undergoes free evolution, which
returns the density matrix into exactly the same state one period later
(neglecting effects of finite $Q$-factor and unharmonicity).
Therefore, the free evolution produces no effect, and measurement pulses
are ``stacked one upon another'', so that the measurement strength adds up,
and many imprecise measurements become equivalent to one very precise
measurement. When the precision of such measurement becomes
better than $\Delta x_0$, the $x$-width (uncertainty) of the resonator
state (density matrix)
necessarily becomes smaller than the ground state width, so the squeezed
state is produced.
Exactly the same mechanism of squeezing works when the measurement pulses
are separated by integer number of oscillation periods (even $n$) or
by an odd number of half-periods (odd $n$) since the free evolution
during a half-period results only in the sign change for position and
momentum.

Notice that even though the measurement squeezes the resonator position $x$,
free evolution between pulses makes it a ``breathing'' mode, so
that $x$-width oscillates in time (with frequency $2\,\omega_0$) becoming
periodically larger and smaller than $\Delta x_0$; correspondingly
the momentum uncertainty of the resonator also oscillates and becomes
squeezed below $\hbar /2\Delta x_0$ periodically.
Because of these oscillations, squeezing is usually considered in the
rotating frame, so what is usually discussed is squeezing of one of two
quadrature amplitudes, which can be translated via free evolution
into the position and momentum at time $t=0$. However, in this paper
we prefer to consider explicitly the oscillating time dependence of
the position and momentum uncertainty, and in this sense we often use
terminology of oscillating in time squeezing of position (or momentum).

        Finite duration of each measurement pulse prevents complete
self-compensation of free evolution and consequently prevents infinite
accumulation of the squeezing degree; instead, squeezing saturates after
initial transient period. Explicit account of finite pulse
duration for continuous measurement by a weakly coupled detector is
one of the main differences between our formalism and the standard analysis
of stroboscopic QND measurements.\cite{Braginsky2,Thorne}

        Finite duration of measurement pulses also leads to a random
motion (diffusion-like) of the wave packet center at the moments of
maximum $x$-squeezing. This can be explained as a consequence of random
momentum kicks during measurement pulses, which are the quantum back-action
price for $x$-measurement. Since the free evolution between the pulses
is not cancelled exactly, momentum kicks lead to gradual $x$-evolution
as well. This effect causes gradual ``heating'' of the nanoresonator.
If not stopped by the damping due to finite $Q$-factor, the resonator
energy will grow up to the effective detector temperature which is on
the order of the detector voltage \cite{Blencowe-rev,MozyrskyMartin}
and is typically very large compared to $\hbar\omega_0$.
This heating can be prevented by using the quantum feedback which
can keep the wavepacket center near zero; such feedback has been
analyzed  by A. Hopkins {\it et al.}, \cite{Hopkins} and we will
basically follow their analysis in the present paper.

        Finite $Q$-factor of the nanoresonator, finite temperature
of the environment, and resonator unharmonicity obviously decrease
the maximum achievable squeezing.
 In this paper we consider the effects of $Q$-factor and
temperature (though for many results they are neglected), but we
do not consider unharmonicity. We also do not analyze explicitly
the use of the squeezed state for the ultrasensitive force detection;
however, we discuss the procedure of squeezed state verification,
which is a closely related topic.
        In the next Section we develop Bayesian formalism for
the analysis of our setup; it is shown to coincide with
the formalism of conditional evolution used in previous papers,
in particular in Refs.\ \onlinecite{DohertyJacobs} and \onlinecite{Hopkins}.
Measurement modulation and simplified equations for Gaussian state
are discussed in Section III,
Section IV is devoted to the calculation of
squeezing, quantum feedback is analyzed in Section V, verification
of the squeezed state is discussed in Section VI, and Section VII is
the conclusion.

\section{Model and Bayesian formalism}

    For simplicity we consider the nanoresonator (Fig.\ \ref{schematic})
measured by the low-transparency QPC (though our results are applicable
to the RF-SET as well), and the system Hamiltonian is
        \begin{equation}
  {\cal H} = {\cal H}_{0} + {\cal H}_{det} + {\cal H}_{int} + {\cal H}_{env}
+ {\cal H}_{fb},
        \label{Hamiltonian}
        \end{equation}
where the first term describes the oscillator:
        \begin{equation}
  {\cal H}_{0} = \frac{\hat{p}^2}{2 m} + \frac{m \omega_0^2 \hat{x}^2}{2}
        \label{oscillator}
        \end{equation}
($\hat{p}$ and $\hat{x}$ being the momentum and position operators),
the last term
        \begin{equation}
  {\cal H}_{fb} = - {\cal F} \hat{x}
        \end{equation}
describes the feedback control of the nanoresonator by applying the force
${\cal F} (t)$, ${\cal H}_{det}$ and ${\cal H}_{int}$ correspond to
the detector and its interaction with the nanoresonator
similar to Refs.\ \onlinecite{Gurvitz} and \onlinecite{MozyrskyMartin}:
        \begin{eqnarray}
   &&
{\cal H}_{det} = \sum\nolimits_l E_l a_l^\dagger a_l +
\sum\nolimits_r E_r a_r^\dagger a_r
      \nonumber \\
 &&\hspace{1.3cm}
+\sum\nolimits_{l,r} ( M \, a_r^\dagger a_l+ \mbox{H.c.}) \, ,
\label{Hdet}
\\
&& {\cal H}_{int}= \sum\nolimits_{l,r} ( \Delta M \, \hat{x} \,
a_r^\dagger a_l + \mbox{H.c.} ) \, ,
        \label{Hint}
        \end{eqnarray}
finally, ${\cal H}_{env}$ describes nanoresonator interaction with phonon
bath at temperature $T$, this interaction is assumed to be weak and
leads to a large quality factor $Q\gg 1$.
In Eqs.\ (\ref{Hdet}) and (\ref{Hint})
 $a^\dagger_{l,r}$ and $a_{l,r}$ are the creation and annihilation
operators for two electrodes of the QPC, for simplicity we assume
no relative phase between the tunneling amplitudes $M$ and $\Delta M$
(taking this phase into account is simple, \cite{Kor-Av,GoanMilburn,Kor-nonid}
but makes the formalism significantly lengthier -- see Appendix).
For a given position $x$ of the oscillator, the average
detector current is $I_x=2\pi |M+\Delta M x|^2 \rho_l\rho_r e^2V/\hbar$,
where $V$ is the QPC voltage which may vary in time with frequency $\omega$
comparable to $\omega_0$, $e$ is the electron
charge, and $\rho_{l,r}$ are the densities of states in the electrodes.

        We assume a weak response of the detector, $|I_x-I_{x'}| \ll
|I_x+I_{x'}|$, and therefore the linear dependence of the detector
current on the measured position
        \begin{equation}
I_x = I_0 + k x,
        \label{linearity}
        \end{equation}
neglecting effects of detector nonlinearity.\cite{Mao}
Also, we neglect the dependence on $x$ of the detector current spectral
density $S_I$ which is assumed to be flat in the frequency range of
interest. Because the voltage $V$ varies in time, $I_0$, $k$, $I_x$,
and $S_I$ also depend on time, that will be taken into account explicitly
in the next Section. Notice that the white noise $S_I$ is an intrinsic
detector noise, which is defined for a fixed voltage on a time scale
much shorter
than the time scale of voltage variations, while the long-time spectral
density of the detector current is obviously affected by the voltage changes
as well as by the oscillating signal from the nanoresonator.

        To describe the dynamics of the continuous quantum measurement
process, we apply the quantum Bayesian approach practically following
the derivation \cite{Kor-99-01} for the case of qubit measurement.
We will have to use similar assumptions; in particular, for the validity
of the Markovian approximation
we assume that the internal dynamics of the detector is much faster
than the oscillator dynamics (this requires $eV \gg \hbar\omega_0$),
we also assume that  detector current is quasicontinuous (which requires
$I_0/e \gg \omega_0$
and even stronger inequality $k\Delta x_0/e \gg \omega_0$).

        To start the derivation, we first neglect the nanoresonator evolution
due to ${\cal H}_0$, ${\cal H}_{env}$, and ${\cal H}_{fb}$ (which will
be added later) and assume constant detector voltage $V$ (variations of $V$,
slow on the time scale of detector dynamics, will be taken into account
later just as a parameter variation). Similar to Ref.\ \onlinecite{Kor-99-01},
the derivation of the Bayesian equations can be done in two ways:
``informational'' and ``microscopic''. Let us start with informational
derivation.

        Since the operator of the QPC current commutes with $\hat{x}$,
the detector current is insensitive to the nondiagonal matrix elements
of the resonator density matrix $\rho (x,x')$ in $x$-representation.
For a measurement duration $\tau$ long enough compared to the detector
time scales $\hbar /eV$ and $e/I_0$ (and short compared to the resonator
evolution due to ${\cal H}_0+{\cal H}_{env}+{\cal H}_{fb}$ so that
it can be neglected), the probability distribution of the noisy detector
current averaged over $\tau$,
$\overline{I} =(1/\tau) \int_0^\tau I(t')\, dt'$,
is given by
        \begin{equation}
P(\overline{I},\tau) = \int P_{x}(\overline{I},\tau) \, \rho(x,x,0) \, dx,
        \label{P(I)}
        \end{equation}
where the third argument of $\rho$ is time and
$P_x(\overline{I},\tau)$ is the probability distribution for $\overline{I}$
in the case of the resonator at position $x$. Since $\tau \gg e/I_0$,
this distribution is Gaussian,
        \begin{equation}
P_x(\overline{I},\tau) =
(2\pi D_I)^{-1/2}  \exp [ -{(\overline{I} - I_x)^2}/{2D_I} ],
        \label{I-Gauss}
        \end{equation}
where
$D_I = S_I/2\tau$ is the variance. Notice that $\overline{I}$ is treated
as a classical variable because detector decoherence time
(which is on the order of $\hbar /eV$) is much shorter than $\tau$.

        Since classical and quantum dynamics are indistinguishable
when nondiagonal matrix elements of $\rho$ cannot affect the
evolution, the diagonal matrix elements of $\rho$ should satisfy the
classical evolution of conditional probability given by the Bayes
formula: \cite{Bayes}
        \begin{equation}
\rho(x,x,\tau) =  \frac{\rho(x,x,0) \, P_x(\overline{I},\tau)}
{\int  \rho(\tilde{x},\tilde{x},0)
P_{\tilde{x}}(\overline{I},\tau)\, d\tilde{x}} \, ,
        \label{Bayes-diag}
        \end{equation}
where ${\overline I}$ is now a particular result of actual measurement
(so the evolution of $\rho$ is conditioned on the measurement result
$\overline{I}$).

        Classical Bayes formula cannot tell us anything about the evolution
of nondiagonal matrix elements $\rho(x,x')$; however, we can use an
obvious limitation
        \begin{equation}
|\rho(x,x',\tau)| \leq \sqrt{\rho(x,x,\tau)\, \rho(x',x',\tau)} .
        \label{basic-inequality}
        \end{equation}
Averaging $\rho (x,x',\tau)$ in this inequality over the measurement
result $\overline{I}$
using the distribution (\ref{P(I)}), we transform this limitation into
        \begin{equation}
|\rho(x,x',\tau )| \leq
\sqrt{\rho(x,x,0)\, \rho(x',x',0)}
\, e^{ -(I_x-I_{x'})^2 \tau /4 S_I }
\label{non-diag-decoh} ,
        \end{equation}
where $\rho$ in this inequality is essentially a density matrix for an
unknown measurement result (in other words not conditioned on $\overline{I}$),
and therefore is a usual ensemble-averaged density matrix.\cite{Leggett}
This result can be compared with the result of conventional ensemble-averaged
approach which says that in the large-$V$ limit the measurement by a
low-transparency QPC leads
to nanoresonator decoherence\cite{MozyrskyMartin} as
$\rho (x,x',\tau )=\rho (x,x',0)\exp [-(I_x-I_{x'})^2 \tau /4 S_I]$
(notice the same exponential dependence).
For a pure initial state, $|\rho (x,x',0)|=\sqrt{\rho (x,x,0)\,
\rho (x',x',0)}$,
both results can be valid simultaneously {\it only if} inequality
(\ref{basic-inequality}) actually reaches its upper bound for {\it each}
measurement result. This means that {\it a pure state of resonator
remains pure in the process of measurement}, similar to the case of
qubit measurement.\cite{Kor-99-01}   Notice that complete absence of
decoherence in a particular realization of the measurement process is because
the QPC is an ideal detector, while for the SET the remaining decoherence rate
would not be zero.\cite{Kor-99-01}

        Combining this result with Eq.\ (\ref{Bayes-diag}), we express it as
        \begin{equation}
\rho(x,x',\tau) =  \frac{\rho(x,x',0)
\sqrt{ P_x(\overline{I},\tau)P_{x'}(\overline{I},\tau)}  }
{\int  \rho(\tilde{x},\tilde{x},0)
P_{\tilde{x}}(\overline{I},\tau)\, d\tilde{x}} \, .
        \label{Bayes-gen}
        \end{equation}
We have neglected the possible phase factor $\exp (i\phi_m )$
because in our model $\phi_m =0$ since in the derivation
$\phi_m$ cannot depend on $\overline{I}$ and there is no phase in
the ensemble-averaged
result.\cite{MozyrskyMartin} The absence of phase $\phi_m$ can also
be proven directly using the microscopic model discussed below.
Nevertheless, nonzero $\phi_m$ can be present in somewhat different
models which include ``asymmetry'' of the detector coupling;
\cite{Kor-Av} an example of such case
is when there is a relative phase
\cite{GoanMilburn,Kor-nonid,Mao} between $M$ and
$\Delta M$ in the Hamiltonian (\ref{Hdet})--(\ref{Hint}) (see
Appendix).

        So far we have proven Eq.\ (\ref{Bayes-gen}) only for
a pure initial state $\rho(x,x',0)$. It is also easy to show its validity
for a mixed state. Representing initial state as
$\rho(0)=\sum_i P_i(0)\rho_i(0)$, where $P_i$ are the probabilities of pure
states $\rho_i(0)$, we apply a ``double-Bayesian'' procedure (as in Ref.\
\onlinecite{Kor-nonid}) in which $P_i(\tau )$ is found via classical Bayes
theorem while each $\rho_i(\tau )$ satisfies the quantum Bayes equation
(\ref{Bayes-gen}). Simple algebra shows that evolution of the mixed density
matrix $\rho$ is still described by Eq.\ (\ref{Bayes-gen}).

        Besides using the ``informational'' approach described above,
Eq.\ (\ref{Bayes-gen}) can also be obtained in a ``microscopic'' way.
Similar to the derivation for the qubit measurement, \cite{Kor-99-01}
the evolution can be divided into the sequence of sufficiently short
segments consisting of ``conventional'' evolution of the nanoresonator
and detector, in which all the detector degrees of freedom are traced over,
except the number $n$ of electrons passed through the detector, so that
the magnitude of interest is the combined density matrix $\rho_n(x,x',t)$.
At the boundaries between the segments we collapse the number $n$
according to the orthodox procedure:\cite{vonNeumann} the probability
of a particular ``realized'' $n=n_0$ is equal to
$\int \rho_{n_0}(x,x,t)\, dx$, and the
corresponding density matrix after collapse is
        \begin{equation}
\rho_{n}(x,x',t+0) =\frac{\rho_{n_0}(x,x',t-0)\, \delta_{n,n_0}}
{\int \rho_{n_0}(\tilde{x},\tilde{x},t-0) \, d\tilde{x}} \, ,
        \end{equation}
where $\delta_{n,n_0}$ is the Kronecker symbol. Applying this sequential
collapse procedure to the conventional evolution of $\rho_n(x,x',t)$
described by Eq.\ (6) of Ref.\ \onlinecite{MozyrskyMartin}, we can
obtain our Eq.\ (\ref{Bayes-gen}) if the resonator evolution
due to ${\cal H}_0+{\cal H}_{env}+{\cal H}_{fb}$ is neglected and
the limit of large detector voltage is assumed.

        The differential equation describing evolution of the resonator
state due to measurement can be obtained by differentiating Eq.\
(\ref{Bayes-gen}) over time $\tau$ at $\tau=0$ and using Eq.\ (\ref{I-Gauss})
(because of the Markovian
approximation, this can be done for arbitrary starting time $t$):
        \begin{eqnarray}
\dot{\rho}(x,x',t) &= &
\rho(x,x',t)\, S_I^{-1}
\{ I(t) [ I_x + I_{x'} - 2\langle I (t) \rangle ]
 \nonumber \\
&& - [ I_x^2 + I_{x'}^2 - 2\langle I^2 (t) \rangle ]/2 \} ,
        \label{dif-eq1}
        \end{eqnarray}
where we have introduced notations
$\langle I (t) \rangle = \int I_x \rho (x,x,t)\, dx$ and
$\langle I^2 (t) \rangle = \int I_x^2 \rho (x,x,t) \, dx$,
while Eq.\ (\ref{P(I)}) transforms into
        \begin{equation}
I(t) = \langle I(t) \rangle + \xi(t),
        \label{noisy-curr}
\end{equation}
where $\xi(t)$ is a white noise with spectral density $S_I$.
   Notice that Eq.\ (\ref{dif-eq1}) actually does not require the current
linearity (\ref{linearity}) and formally
coincides with the similar equation for the case of
entangled qubits measured by an ideal detector, \cite{Kor-nonid}
if $x$ is replaced by the index corresponding to the state of qubits.
For the linear detector with response (\ref{linearity}), Eq.\
(\ref{dif-eq1}) becomes
        \begin{eqnarray}
\dot{\rho}(x,x') &= &
\rho(x,x')\, S_I^{-1}
\{ [I(t)-I_0]k ( x + x' - 2\langle x \rangle )
 \nonumber \\
&& - (k^2/2)[ x^2 + (x')^2 - 2\langle x^2 \rangle ] \} ,
        \label{dif-eq2}
        \end{eqnarray}
where for brevity we do not show explicitly the time dependence of $\rho$,
$\langle x\rangle =\int x \rho(x,x)\, dx$, and
$\langle x^2\rangle =\int x^2 \rho(x,x)\, dx$.

  Notice that Eq.\ (\ref{dif-eq1}) have been obtained by differentiating
Eq.\ (\ref{Bayes-gen}) over $\tau$ using the standard rules (i.e.\
in the first order). Therefore  Eqs.\ (\ref{dif-eq1}) and (\ref{dif-eq2})
are the stochastic equations in the so-called Stratonovich form which
assumes ``centered'' definition of the derivative, $\dot\rho (t) \equiv
\lim_{\tau\rightarrow 0} [\rho (t+\tau /2)- \rho (t-\tau /2)]/\tau$,
and allows us to use standard calculus rules. \cite{Oksendal}
For a nonlinear stochastic equation the calculus rules are quite different
for another widely used definition of the ``forward'' derivative,
$\dot\rho (t) \equiv \lim_{\tau\rightarrow 0} [\rho (t+\tau )-
\rho (t)]/\tau$, which would lead to  an equation in the so-called It\^o
form. \cite{Oksendal}
Advantage of the It\^o form is the simple averaging over the noise
(while averaging in Stratonovich form is not trivial); this is the reason
why It\^o form is usually preferred by mathematicians, even though
physical intuition works better in the Stratonovich form. Translation
back and forth between two forms is often useful to solve a particular
problem.

        The rule of translation between the two forms is the following:
\cite{Oksendal} for a system of equations $\dot{y}_i(t)= G_i({\bf y},t)+
F_i({\bf y},t) \xi(t)$ in the Stratonovich form, the corresponding
It\^o equation is
$\dot{y}_i(t)= G_i({\bf y},t)+ F_i({\bf y},t) \xi(t)+
(S_\xi /4) \sum_j [d F_i({\bf y},t) /d y_j] F_j ({\bf y},t)$,
where $y_i$ are the components of the vector ${\bf y}$, $G_i$ and $F_i$
are arbitrary functions, and $S_\xi$ is the spectral density of white
noise $\xi (t)$. To apply this rule to our case (${\bf y}=\rho$),
we replace index $i$ by continuous set $(x,x')$ and replace summation by
integration; then Eq.\ (\ref{dif-eq2}) is translated into the It\^o form as
        \begin{eqnarray}
\dot{\rho}(x,x') & = &
 (k/S_I)\, (x+x'-2\langle x\rangle ) \, \rho(x,x') \, \xi(t)
 \nonumber \\
&&  - (k^2/4S_I)\, (x-x')^2 \, \rho(x,x') .
        \label{dif-eq3}
        \end{eqnarray}
This equation is similar to equations derived in many publications
(e.g., in Refs.\ \cite{Mensky,GWM,DohertyJacobs,Hopkins})
for measurement of a mechanical oscillator.
Notice that the last term in Eq.\ (\ref{dif-eq3}) does not describe
decoherence
in a particular realization of the measurement (recall that a pure state
remains pure); however, it describes ensemble decoherence, since averaging
over the measurement result (over noise $\xi$) is done in It\^o form simply
by using $\xi =0$. This term can also be
rewritten in a standard double-commutator form (see, e.g., Refs.\
\cite{Lindblad,Leggett,CavesMilburn,GWM,ZurekHabibPaz-prl,MozyrskyMartin,%
DohertyJacobs,Hopkins})
since $(x-x')^2\rho (x,x') = [\hat{x}, [\hat{x},\rho]]_{x,x'}$.

        Equation (\ref{dif-eq3}) describes the evolution of the nanoresonator
state due to measurement by the ideal detector (low-transparency QPC)
described by the Hamiltonian (\ref{Hdet})--(\ref{Hint}). To extend the
formalism to a nonideal detector (for example, RF-SET) we introduce its
quantum efficiency (ideality) $\eta \leq 1$  similar to Ref.\
\onlinecite{Kor-99-01}
and replace the decoherence factor $k^2/4S_I$ by $k^2/4S_I\eta$.
Simply speaking, $1/\eta$ is the ratio between the product of output and
back-action noises of the detector and its quantum-limited value.
\cite{Kor-99-01,Averin-eta,Clerk-pd} For example, $k^2/4S_I$ in Eq.\
(\ref{dif-eq3}) can be replaced by $k^2/4S_I\eta$ when an extra term
$-\gamma_{cl} (x-x')^2\rho(x,x')$ is due to additional classical back-action
noise from the detector or when the output noise of the detector
contains an additional noise (see Ref.\ \onlinecite{Kor-nonid} and Appendix).

        As the final step of our derivation, we add into Eq.\ (\ref{dif-eq3})
(modified by efficiency $\eta$) the evolution due to terms
${\cal H}_0+{\cal H}_{evn}+{\cal H}_{fb}$ of the Hamiltonian
(\ref{Hamiltonian}). Interaction with the thermal bath denoted by
${\cal H}_{env}$ can be described by the standard Brownian motion master
equation. \cite{Gardiner} Assuming weak coupling (large $Q$-factor)
and arbitrary temperature $T$, we add damping and diffusion terms
\cite{Caldeira-Caldeira,Hopkins}
$-(\imat\omega_0/2\hbar Q) [ \hat{x}, \{\hat{p},\rho\}_+]
-(m \omega_0^2/2 \hbar Q)\,\coth (\hbar \omega_0/2 T) \,
     [ \hat{x},[\hat{x},\rho ]]$
into the equation for $\dot \rho$. Therefore, our final equation
for the nanoresonator evolution is (in It\^o form)
        \begin{eqnarray}
&& \hspace{-.5cm}
\dot{\rho}(x,x') = \frac{- i}{\hbar} [{\cal H}_0+{\cal H}_{fb},\rho ]_{x,x'}
-\frac{\imat \omega_0}{2 \hbar Q} \, [ \hat{x}, \{\hat{p}, \rho\}_+
]_{x,x'}
\nonumber\\
&& \hspace{-0.1cm}
 -\left( \frac{k^2}{4S_I \eta}
+\frac{m \omega_0^2 }{2 \hbar Q}\,\coth{\frac{\hbar \omega_0}{2 T}}
+\gamma_{add}\right) (x-x')^2 \, \rho(x,x')
\nonumber\\
&& \hspace{-0.1cm}
+ \frac{k}{S_I} \, (x+x'-2\langle x\rangle ) \, \rho(x,x') \, \xi(t)
,
\label{meas-Ito-linear}
        \end{eqnarray}
in which the white noise $\xi (t)$ is related to the detector current $I(t)$
via Eq.\ (\ref{noisy-curr}) as $\xi (t) =I(t) I_0 - k\langle x (t)\rangle$
 and $\gamma_{add}$ is introduced
phenomenologically to take into account sources of additional dephasing,
for example due to high-temperature electromagnetic fields penetrating
into cryostat (we will mostly assume $\gamma_{add}=0$).
Notice that there is no damping term due to measurement (in contrast to
the term due to $Q$-factor) because we treat detector as a device with
completely classical output and therefore detector voltage is very large
while its unnormalized coupling is very weak; in other words, we assume that
the nanoresonator energy is limited  well below the
effective temperature of detector (which is on the order of $eV$)
by other effects (feedback and $Q$-factor).

        \section{Measurement modulation and equations for Gaussian states}

        Periodic modulation of the QPC voltage $V = f(t) V_0$ (with
frequency comparable to $\omega_0$ and much smaller than $eV/\hbar$ and
$I_0/e$) leads to the corresponding modulation
of the measurement parameters:
$k=f(t)k_0$, $I_x= f(t) (I_{00}+k_0 x)$,
$S_I =|f(t)| S_0$,
so that the ``measurement strength'' $k^2/S_I$ is modulated as
$|f(t)|k_0^2/S_0$ [in general $f(t)$ may be negative].
In the case of RF-SET the voltage dependence of
parameters is not trivial (also the modulation can be arranged using
the gate voltage instead of the bias), but we still can define $f(t)$ as
the modulation of the measurement strength from equation
        \begin{equation}
 k^2/S_I = |f(t)| k_0^2/S_0.
        \label{f(t)-def}
        \end{equation}
Quantum efficiency $\eta$ in general can also be affected by the modulation,
but for simplicity we assume it to be constant.

Notice that the noise $\xi (t)$ in Eq.\ (\ref{meas-Ito-linear}) has implicit
time dependence because of modulated in time spectral density $S_I$.
To remove this dependence we define the white noise
        \begin{equation}
\xi_0(t)=\xi (t)  \sqrt{S_0/S_I} \, \mbox{sgn} [f(t)]
        \end{equation}
with time-independent spectral density
$S_0$. Then the last term in Eq.\ (\ref{meas-Ito-linear}) can be written
as $\sqrt{|f(t)|} (k_0/S_0) (x+x'-2\langle x\rangle)\,\rho (x,x')\,\xi_0(t)$.

        Somewhat similar to the case of qubit measurement,
\cite{Kor-99-01}
we define the dimensionless (time-dependent) coupling as
        \begin{equation}
{\cal C}=\frac{\hbar k^2}{S_I m \omega_0^2} =|f(t)| {\cal C}_0 ,
        \label{coupling}
        \end{equation}
which can also be expressed as ${\cal C}=4/\omega_0\tau_m$, where
$\tau_m = 2S_I/(k \Delta x_0)^2$ is the ``measurement'' time which would
be necessary
to distinguish (with signal-to-noise ratio of 1) two position states
separated by the ground state width $\Delta x_0$.
We will mainly consider the case of weak coupling, ${\cal C}\ll 1$,
which corresponds to a realistic experimental situation. As an example,
${\cal C}$ is on the order of $10^{-6}$ for the experimental parameters
of Ref.\ \onlinecite{Schwab-QL}.

        In this paper we will consider two types of modulation with frequency
$\omega$: harmonic modulation with the relative modulation depth
$A_{mod}=(f_{max}-f_{min})/f_{max}$, $0 \leq A_{mod} \leq 2$:

        \begin{equation}
f(t)= 1 + \frac{A_{mod}}{2} (-1+\cos \omega t)
        \label{harm-mod}
        \end{equation}
and the square-wave (stroboscopic) modulation with pulse width $\delta t$
and relative depth $A_{mod}$:
        \begin{equation}
f(t) = \left\{ \begin{array}{l}  1, \,\,\,
|t-j\times 2\pi /\omega |\leq \delta t/2, \,\, j=1, 2, \ldots
\\ 1-A_{mod}, \,\,\, \mbox{otherwise} \, .
        \end{array} \right.
        \label{strob-mod}
        \end{equation}
Notice that $|f(t)|\leq 1$, so ${\cal C}_0$ corresponds to the
maximum coupling.
        Since $f(t)$ reaches zero in both types of modulation at
$A_{mod}\geq 1$ (we will mostly consider 100\% modulation, $A_{mod}=1$),
the conditions
$eV\gg \hbar \omega_0$ and $k\Delta x_0/e\gg \omega_0$ required for the
Bayesian formalism are violated during a fraction of the modulation period.
However, the expected corrections to the Bayesian equations (see, e.g.\
Ref.\ \onlinecite{MozyrskyMartin}) have the relative strength of crudely
$\hbar\omega_0/eV$, which means that the poorly-described evolution during
these fractions of the period is quite slow. Therefore, we can still
use Eq.\ (\ref{meas-Ito-linear}) for the analysis in the case of
sufficiently large maximum voltage, when the neglected contribution to the
evolution during low-voltage phase is significantly smaller than the
well-described contribution during large-voltage phase.
    The neglected contribution is expected to lead to a weak relaxation
of the nanoresonator state and can crudely be taken into account
as some reduction of the $Q$-factor.

Following Refs.\ \cite{HalliwellZoupas,BreslinMilburn,DohertyJacobs,%
Hopkins}, we assume that the oscillator state can be described as a
Gaussian state. This assumption can be justified by the fact that
a Gaussian state remains Gaussian in the process of continuous
measurement \cite{BreslinMilburn} (we have checked this statement for
nonideal detectors including ``asymmetric'' detectors and for varying
in time strength of measurement) and by the fact that the thermal state
(natural initial condition) is Gaussian. \cite{Gardiner} It is also
known \cite{HalliwellZoupas} that any initial pure state approaches
a Gaussian state in a course of continuous measurement by an ideal detector.
We have also checked that a mixture of Gaussian states evolves into
a single Gaussian state due to measurement.

        A Gaussian state is defined \cite{Gardiner} as a state for which
the Wigner function
$W(x,p) \equiv (\pi\hbar)^{-1} \int \rho(x+x', x-x')\, \exp (-2ix'p/\hbar)
\, dx'$ has a Gaussian form:
        \begin{eqnarray}
&& W(x,p) = Norm \times \exp \left( -  {\bf B}^T {\bf D}^{-1} {\bf B} /2
 \right) ,
        \nonumber \\
&& {\bf B}=\left( \begin{array}{c}   x-\langle x\rangle \\
                            p-\langle p\rangle
        \end{array} \right), \hspace{0.3cm}
{\bf D} = \left( \begin{array}{cc}     D_x & D_{xp} \\
                                     D_{xp} & D_p
               \end{array} \right)  ,
        \nonumber
        \end{eqnarray}
with normalization factor $Norm =[2\pi (D_x D_p-D_{xp}^2)^{1/2}]^{-1}$.
In $x$-representation the density matrix of this state is
\begin{eqnarray}
&& \hspace{-0.9cm}
\rho(x,x')=\frac{1}{\sqrt{2\pi D_x}} \,
\exp{\left[-\frac{(\frac{x+x'}{2}-\langle x\rangle )^2}{2 D_x}\right]}
\nonumber\\
&& \hspace{-0.7cm}
\times \exp{\left[-\frac{(x-x')^2}{8 D_x} \,
\frac{(D_x D_p - D_{xp}^2)}{\hbar^2/4} \right]}
\nonumber\\
&& \hspace{-0.7cm}
\times \exp{\left[ i (x-x') \left( \frac{\langle p\rangle }{\hbar} +
\left( \frac{x+x'}{2}-\langle x\rangle \right)
\frac{D_{xp}}{\hbar D_x} \right) \right]} .
        \label{Gaussian-den-mat}
        \end{eqnarray}

Gaussian state is characterized by only five real parameters:
average position $\langle x\rangle = \langle \hat{x}\rangle$ and
momentum $\langle p\rangle =\langle \hat{p}\rangle$,
their variances $D_x=\langle \hat{x}^2\rangle -\langle \hat{x}\rangle^2$
and $D_p=\langle \hat{p}^2\rangle -\langle \hat{p}\rangle^2$,
and the correlation $D_{xp} =\langle \hat{x}\hat{p}+\hat{p}\hat{x}\rangle /2
-\langle\hat{x}\rangle \langle\hat{p}\rangle$. These parameters satisfy
the generalized Heisenberg inequality \cite{Man'ko}
        \begin{equation}
D_x D_p - D_{xp}^2 \geq \hbar^2/4 ,
        \label{uncertain}
        \end{equation}
which reaches the lower bound for the pure states. In particular,
``coherent'' states are the Gaussian states with $D_x=(\Delta x_0)^2$,
$D_p=(\hbar /2\Delta x_0)^2$, and $D_{xp}=0$.

        For Gaussian states Eq.\ (\ref{meas-Ito-linear}) significantly
simplifies and transforms into the following set of equations:
        \begin{eqnarray}
&& \hspace{-1.0cm}
\dot{\langle x\rangle } =\frac{\langle p\rangle}{m}
+ \frac{2k_0}{S_0}\, |f(t)|^{1/2}\, D_x\, \xi_0(t) \, ,
\label{eq-m-x}\\
&& \hspace{-1.0cm}
\dot{\langle p\rangle }=-m \omega_0^2 \langle x\rangle
+ \frac{2k_0}{S_0} |f(t)|^{1/2} D_{xp} \xi_0(t)
-\frac{\omega_0}{Q}\langle p\rangle +{\cal F}  ,
\label{eq-m-p}\\
&& \hspace{-1.0cm}
\dot{D}_{x}=\frac{2}{m}\, D_{xp}
- \frac{2k_0^2}{S_0}\, |f(t)|\, D_x^2 \, ,
\label{eq-mDx}\\
&& \hspace{-1.0cm}
\dot{D}_{p}=-2m \omega_0^2 D_{xp} +
\frac{k_0^2 \hbar^2}{2S_0 \eta}\ |f(t)|- \frac{2k_0^2}{S_0}\ |f(t)|\,
        D_{xp}^2
        \nonumber \\
&& \hspace{0.0cm}
-\frac{2\omega_0}{Q}\, D_p +\frac{\hbar m\omega_0^2}{Q}
\, \coth{\frac{\hbar \omega_0}{2T}} + 2\hbar^2\gamma_{add} \, ,
\label{eq-mDp}\\
&& \hspace{-1.0cm}
\dot{D}_{xp}=\frac{D_p}{m}  -m \omega_0^2 D_{x}-
\frac{2k_0^2}{S_0}\, |f(t)|\, D_x D_{xp} -\frac{\omega_0}{Q}\, D_{xp}  ,
\label{eq-mDxp}
\end{eqnarray}
        which practically coincide with the equations derived in Refs.\
\onlinecite{DohertyJacobs} and \onlinecite{Hopkins}
(see also Ref.\ \onlinecite{HalliwellZoupas}),
except for the time dependent
$f(t)$. It is interesting to notice that while Eq.\ (\ref{meas-Ito-linear})
is a nonlinear stochastic equation, for which the Stratonovich and Ito
forms are significantly different, there is no such difference for
Eqs.\ (\ref{eq-m-x})--(\ref{eq-mDxp}), so they can be treated as simple
ordinary differential equations.

        Notice that the equations for $D_x$, $D_p$, and $D_{xp}$ do not
depend on noise $\xi_0 (t)$ and feedback force ${\cal F}$, and are
decoupled from the remaining equations.
Therefore the evolution of the ``wavepacket width'' $\sqrt{D_x}$ is
deterministic. Analyzing the possibility to squeeze the nanoresonator state,
we will consider separately squeezing of the variance $D_x$
and contribution $D_{\langle x\rangle}$
due to fluctuating position of the packet
center $\langle x\rangle$. As will be discussed in the next Section,
$D_x$ may be made significantly smaller than the ground state variance
$\Delta x_0^2$  using modulation $f(t)$,
while in Section V we show that $D_{\langle x\rangle}$ can be made even
smaller using the feedback.

        \section{Wavepacket width squeezing}

        In this Section we analyze Eqs.\ (\ref{eq-mDx})--(\ref{eq-mDxp})
and show that the $x$-width $\sqrt{D_x}$ of the nanoresonator state
can be made much smaller than $\Delta x_0=\sqrt{\hbar /2m\omega_0}$.
      Let us use the natural normalization of $D_x$ and $D_p$ by the ground
state parameters, $d_x\equiv D_x/(\hbar/2m\omega_0)$,
$d_p\equiv D_p/(\hbar m\omega_0/2)$, and similarly
$d_{xp}\equiv D_{xp}/(\hbar/2)$. Then Eqs.\ (\ref{eq-mDx})--(\ref{eq-mDxp})
can be rewritten as
        \begin{eqnarray}
&& \hspace{-0.5cm}
\dot{d}_{x}/\omega_0 =  2 d_{xp}
- {\cal C}_0\,|f(t)|\, d_x^2 \, ,
\label{eq-mDx-dles}\\
&& \hspace{-0.5cm}
\dot{d}_{p}/\omega_0 = \ -2 d_{xp} +
({\cal C}_0/\eta) \,  |f(t)|- {\cal C}_0 \, |f(t)|\, d_{xp}^2
  \nonumber \\
&& \hspace{0.9cm}
-\frac{2}{Q}\, d_p +  \frac{2}{Q}\,\coth{\frac{\hbar\omega_0}{2T}}
+\frac{4\hbar\gamma_{add}}{m\omega_0^2}    \, ,
\label{eq-mDp-dles}\\
&& \hspace{-0.5cm}
\dot{d}_{xp}/\omega_0 =  d_{p} -d_{x}-
{\cal C}_0 \, |f(t)|\, d_x d_{xp} -\frac{1}{Q}\, d_{xp} \, .
\label{eq-mDxp-dles}
\end{eqnarray}

        It is easy to see that the effect of additional dephasing
$\gamma_{add}$ is equivalent (in case of finite $Q$-factor) to increase
of environment temperature $T$; so we will not consider this effect
separately ($\gamma_{add}=0$ is assumed in the rest of the paper).
Also, let us postpone the analysis of effects due to finite $Q$ and
temperature
until Subsection D and start with the case of infinite $Q$-factor.

        \subsection{Numerical results for squeezing degree
${\cal S}$}

\begin{figure}
\centering
\includegraphics[width=2.7in]{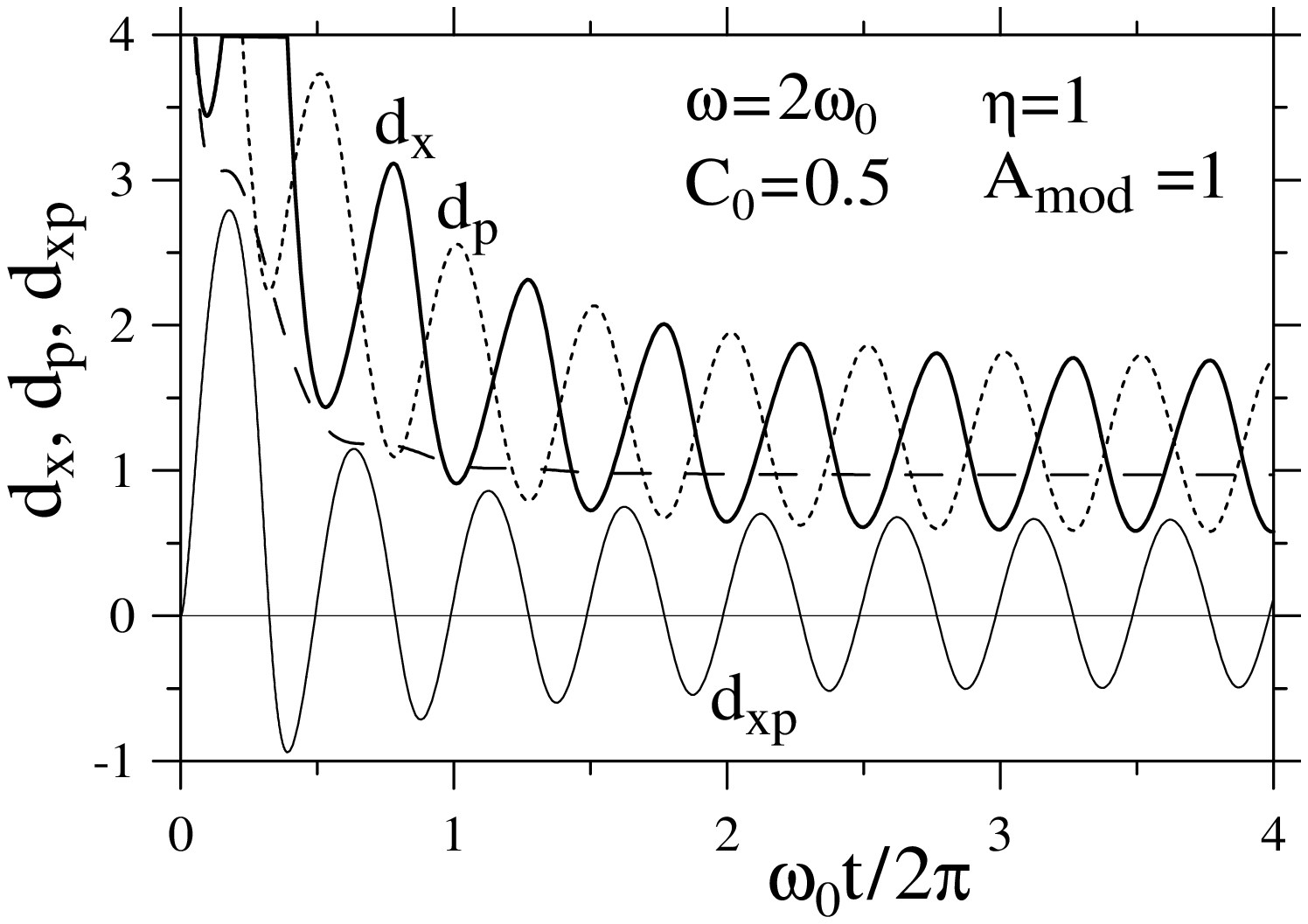}
\includegraphics[width=2.7in]{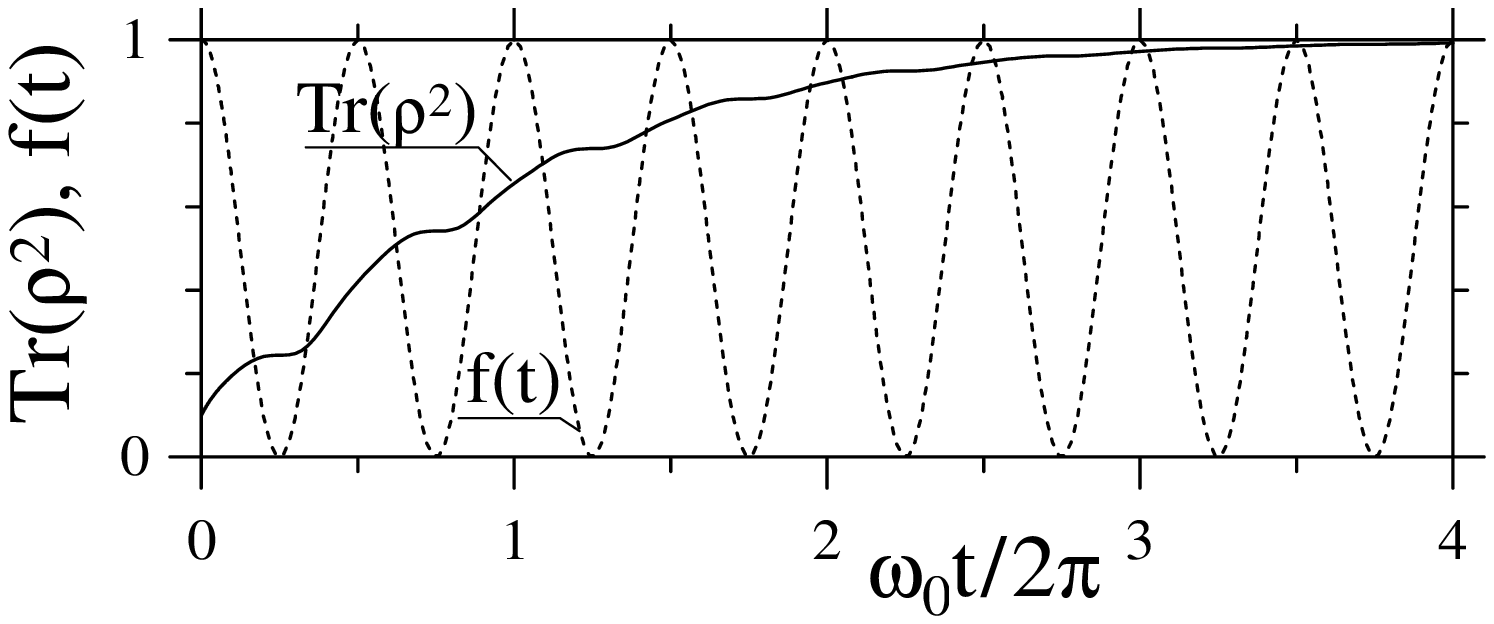}
% where an .eps filename suffix will be assumed under latex,
% and a .pdf suffix will be assumed for pdflatex
\caption{
An example of time dependence of dimensionless wavepacket variances
$d_x$, $d_p$, and $d_{xp}$ (upper panel) for harmonic modulation
$f(t)$ of measurement strength shown in the lower panel. After a transient
period the evolution reaches stationary oscillating regime. The state
purity $\mbox{Tr}(\rho^2)$ (lower panel) gradually approaches unity
(mixed initial state with $d_x=d_p=10$ is chosen).
Long-dashed line in the upper panel shows evolution of $d_x$ in the
non-modulated case $f(t)=1$.
}
\label{fig-time}
\end{figure}

        We have analyzed Eqs.\ (\ref{eq-mDx-dles})--(\ref{eq-mDxp-dles})
numerically for the harmonic
(\ref{harm-mod}) and stroboscopic (\ref{strob-mod}) modulation $f(t)$
for several values of the maximum coupling ${\cal C}_0$, concentrating on
the range ${\cal C}_0\lesssim 1$.
Notice that for the stroboscopic modulation the evolution during each
period of modulation can be calculated analytically using Riccati equations
\cite{DohertyJacobs} that significantly simplifies the numerical
calculations.
 As anticipated, we have found that irrespectively
of the initial conditions, Eqs.\ (\ref{eq-mDx-dles})--(\ref{eq-mDxp-dles})
approach the asymptotic solutions which oscillate with the modulation
frequency $\omega$ (Fig.\ \ref{fig-time}).
Even for small coupling, ${\cal C}_0 \ll 1$, the
asymptotic oscillations can be significant in the case of resonance:
$\omega \simeq 2\,\omega_0 /n$ (notice that
at ${\cal C}_0=0$ the variances oscillate with frequency $2\,\omega_0$).
During the oscillation period the asymptotic solution for $d_x (t)$ reaches
the values
both above and below the stationary solution for $f(t)=1$ which
is \cite{DohertyJacobs,Hopkins}
        \begin{equation}
d_x=(\sqrt{2}/{\cal C}_0) [(1+{\cal C}_0^2/\eta)^{1/2} -1]^{1/2}
        \label{d-stat}
        \end{equation}
 and becomes $d_x=1/\sqrt{\eta}$
for ${\cal C}_0\ll 1$.
    Most importantly, the squeezed state, $d_x < 1$, can be achieved
for both harmonic and stroboscopic modulation.

\begin{figure}
\centering
\includegraphics[width=2.7in]{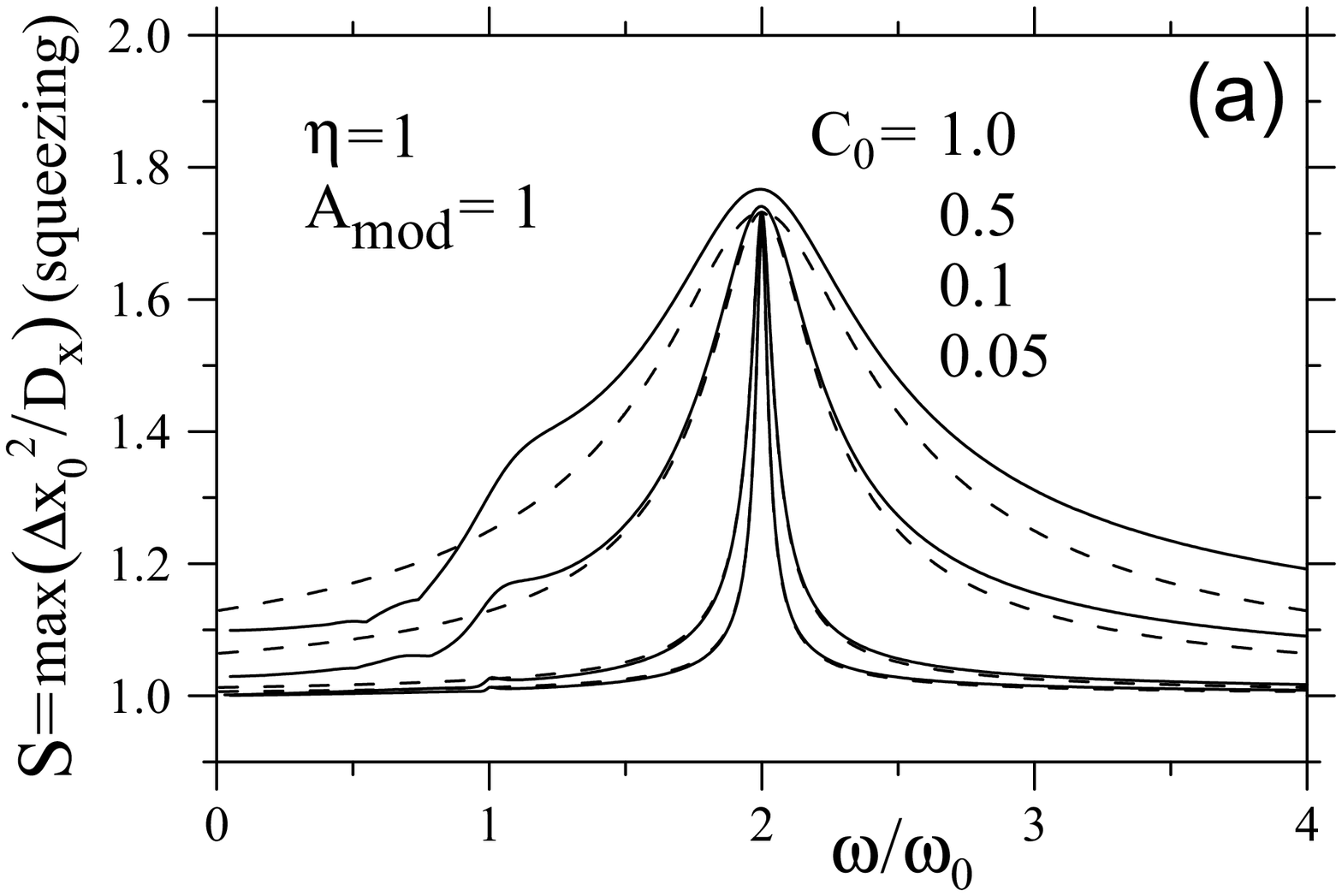}
\includegraphics[width=2.7in]{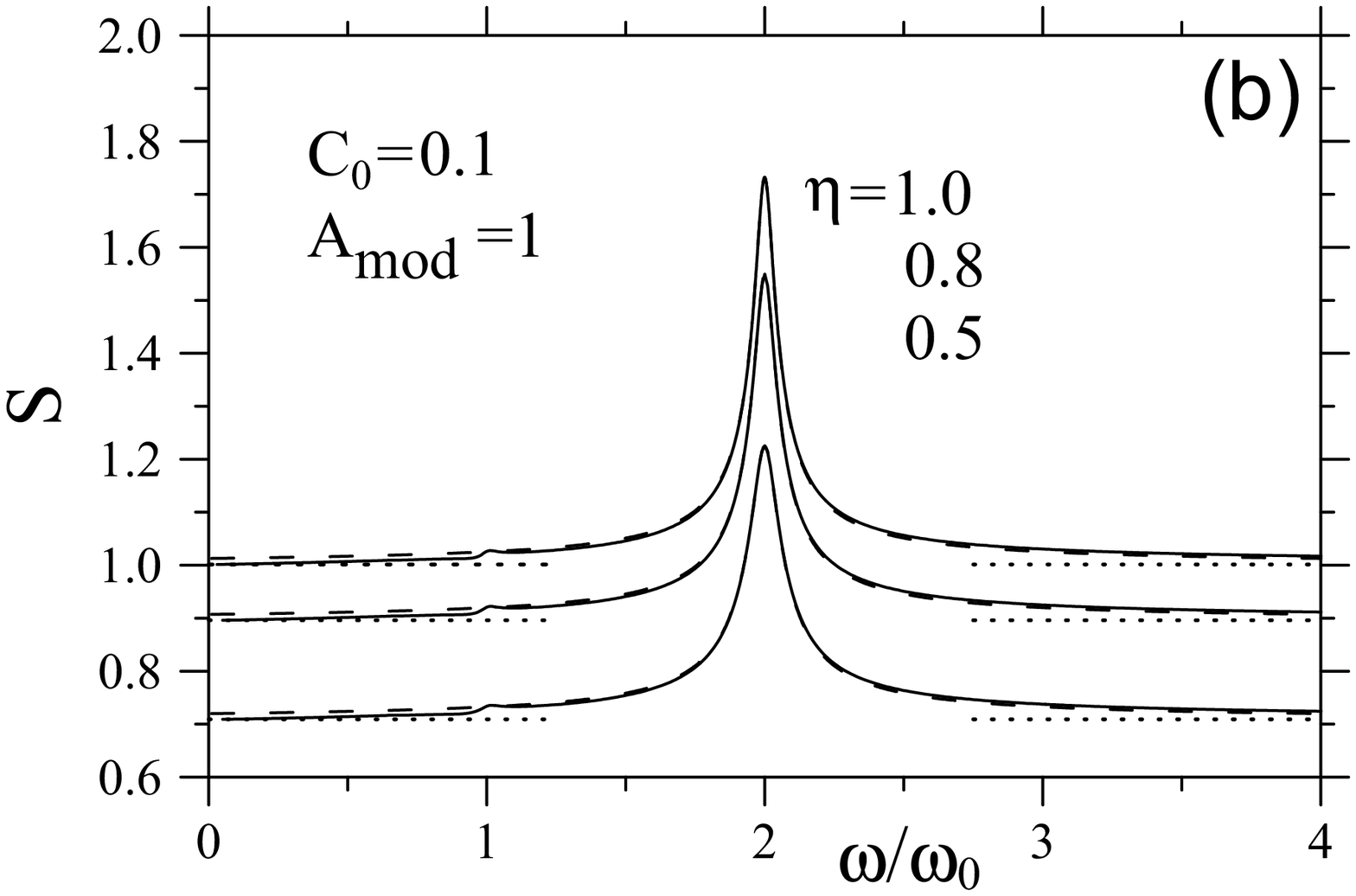}
\includegraphics[width=2.7in]{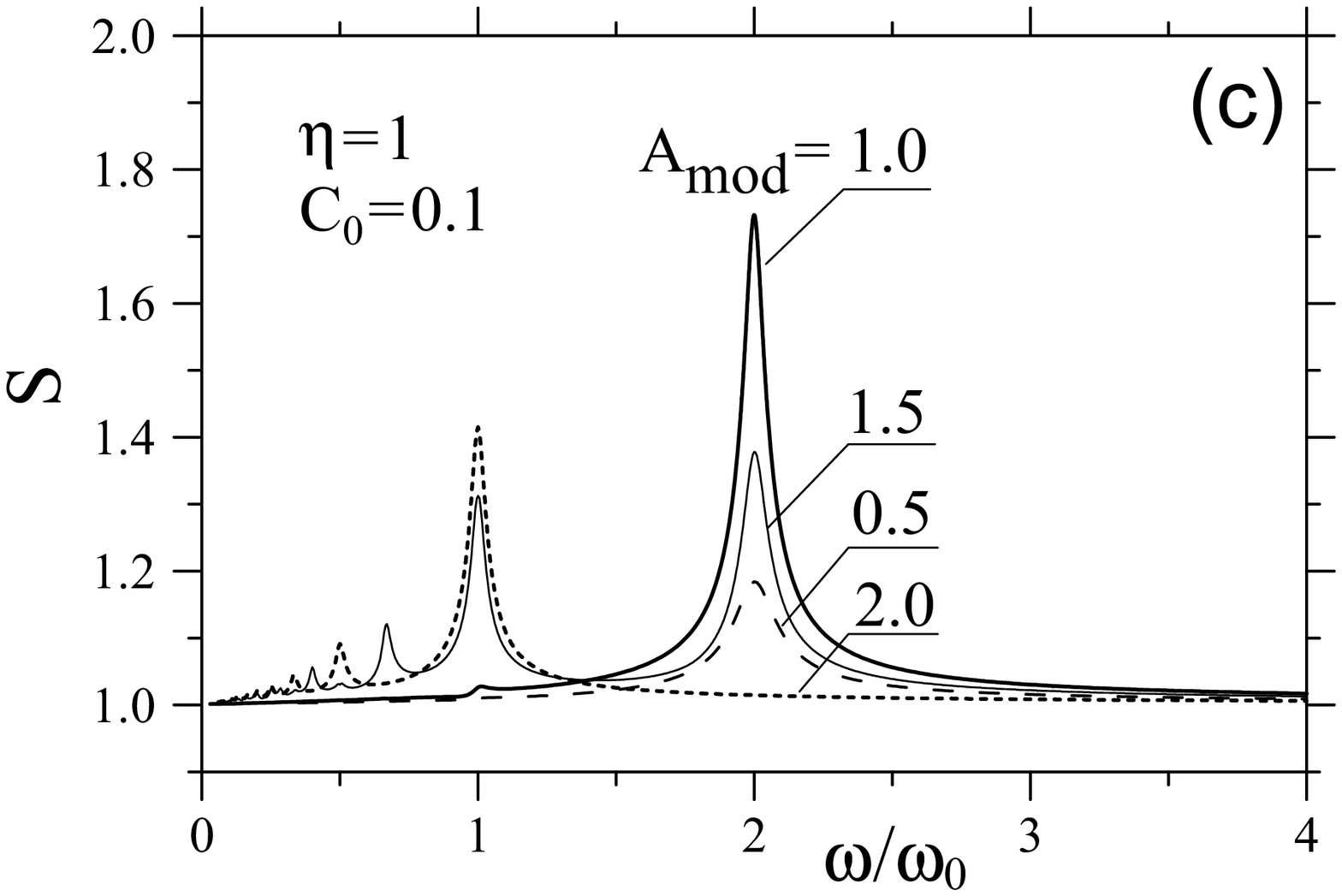}
% where an .eps filename suffix will be assumed under latex,
% and a .pdf suffix will be assumed for pdflatex
\caption{Dependence of the packet width squeezing ${\cal S}$
(maximized over the modulation period) on the frequency $\omega$
of the harmonic modulation (\protect\ref{harm-mod}) of the measurement
strength, for several values of (a) the coupling ${\cal C}_0$,
(b) detector quantum efficiency $\eta$, and (c) modulation amplitude
$A_{mod}$. Solid lines in (a) and (b) are the numerical results
while dashed lines are the analytical results corresponding to
Eqs.\ (\protect\ref{S-A}) and
(\protect\ref{Sq-A-cos}); the dotted lines in (b) are the asymptotes
${\cal S}=\sqrt{\eta}$.
}
\label{fig-cos}
\end{figure}

        Figure \ref{fig-cos} shows the $x$-squeezing maximized over the
oscillation period
for the asymptotic solution, ${\cal S}=\mbox{max}_t [1/d_x(t)] =
\mbox{max}_t [\Delta x_0^2/D_x(t)]$, as a function of the modulation
frequency $\omega$ for the harmonic modulation (\ref{harm-mod})
and several values of coupling ${\cal C}_0$, efficiency $\eta$ and
modulation amplitude $A_{mod}$.
(Notice that in the rotating frame the squeezing ${\cal S}$ does not depend
on time for weak coupling.)
One can see that maximum squeezing is achieved for modulation with
twice the resonator frequency, $\omega =2\,\omega_0$, 100\% amplitude,
$A_{mod}=1$, and for ideal detector, $\eta =1$. The value of maximum
squeezing does not depend much on coupling ${\cal C}_0$
[Fig.\ \ref{fig-cos}(a)] and is equal to
${\cal S}\approx 1.73$ for weak coupling, while the width of resonance
scales proportionally to ${\cal C}_0$ (analytical results discussed later
and shown by dashed lines confirm this behavior).
For nonideal detectors, $\eta <1$ [Fig.\ \ref{fig-cos}(b)],
the height of the peak decreases,
${\cal S}(2\omega_0)\approx 1.73 \sqrt{\eta}$,
and its width increases. Away from the resonance ${\cal S}$ approaches
the value for non-modulated measurement given by Eq.\ (\ref{d-stat})
[dotted lines in Fig.\ \ref{fig-cos}(b)].
Besides the main resonance, there are resonances at $\omega =2\,\omega_0/n$,
$n\geq 2$, which are barely visible in Figs.\ \ref{fig-cos}(a) and
\ref{fig-cos}(b)
and lead to small shoulders rather than to peaks. However, these
resonances become much better visible for modulation amplitudes $A_{mod}$
greater than 100\% as shown in Fig.\ \ref{fig-cos}(c). In particular,
for $A_{mod}=2$ there is no peak at $\omega =2\,\omega_0$, and the main peak
is at $\omega =\omega_0$; this is obviously because
in this case $|f(t)|$ oscillates with frequency $2\,\omega$ instead of
$\omega$.

\begin{figure}
\centering
\includegraphics[width=2.7in]{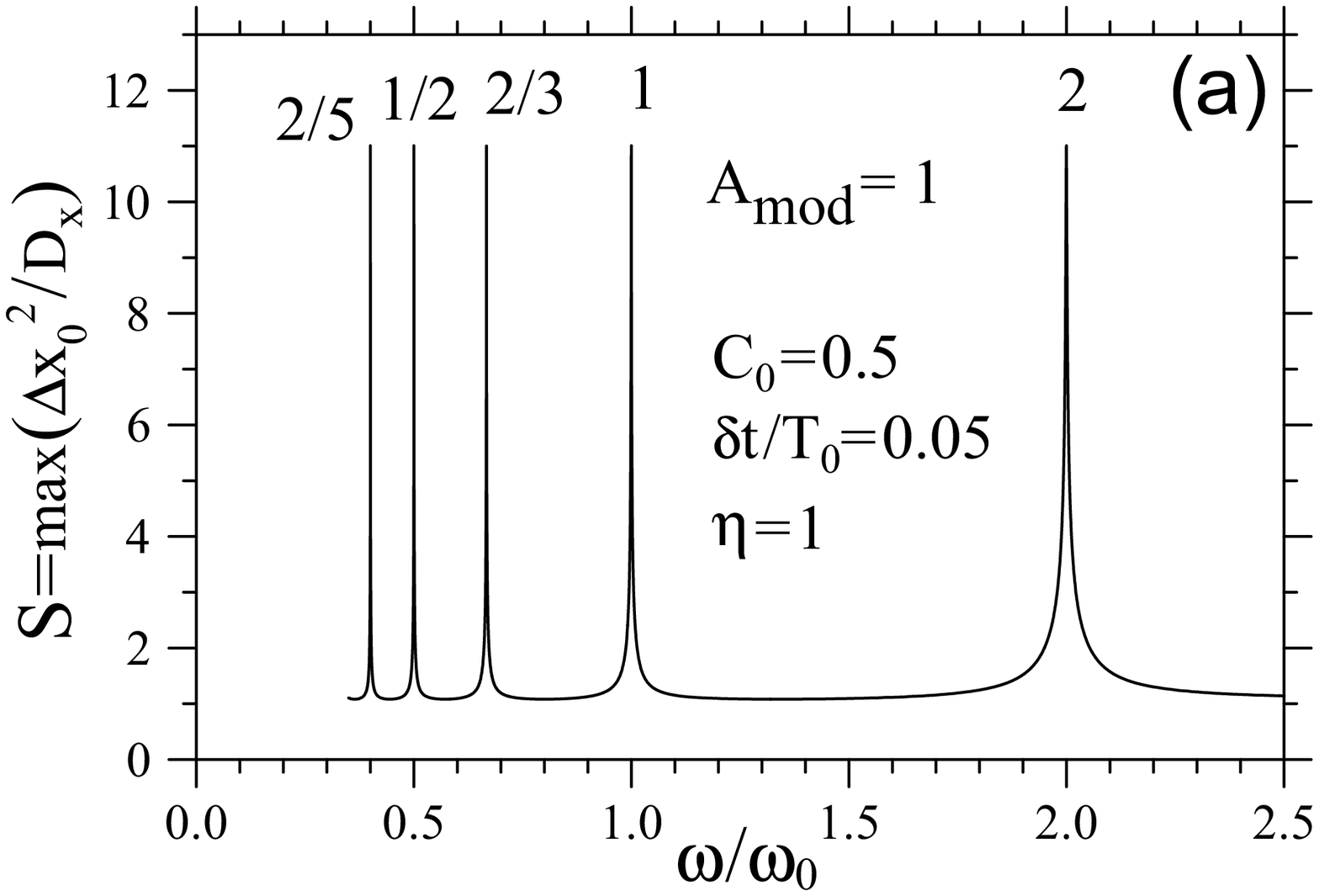}
\includegraphics[width=2.7in]{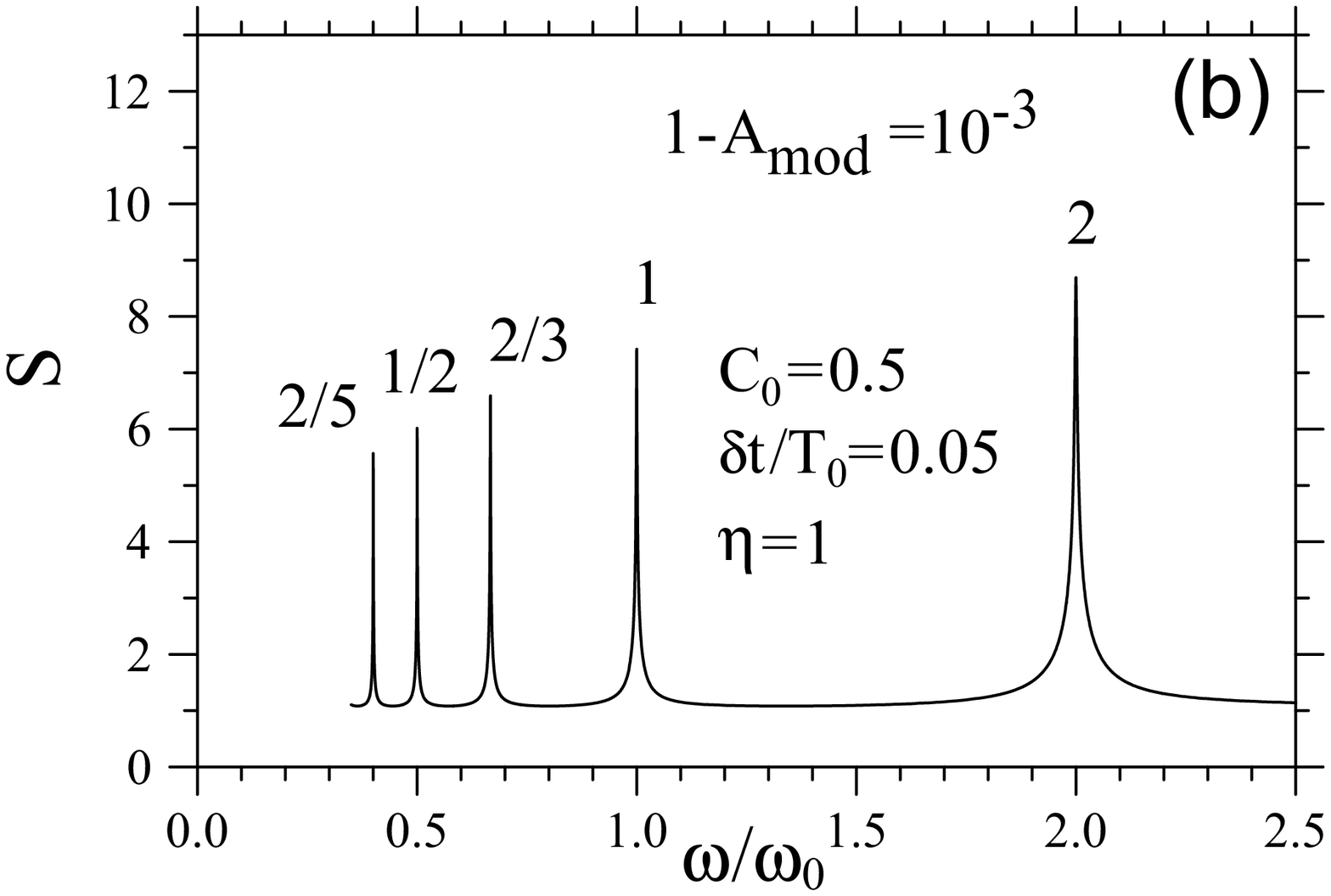}
% where an .eps filename suffix will be assumed under latex,
% and a .pdf suffix will be assumed for pdflatex
\caption{Numerical results for the packet width squeezing ${\cal S}$
as a function of modulation frequency $\omega$ for the stroboscopic
measurement modulation (\protect\ref{strob-mod}) with finite pulse duration
$\delta t$. Efficient squeezing occurs at $\omega \approx 2\,\omega_0/n$.
Infinitely large $Q$-factor of the nanoresonator is assumed.
}
\label{fig-strob1}
\end{figure}

        Much stronger squeezing can be achieved
for the stroboscopic modulation (\ref{strob-mod}) of the measurement.
 Figure \ref{fig-strob1} shows
${\cal S}(\omega )$ for the ideal detector with ${\cal C}_0=0.5$
and pulse duration $\delta t=0.05 \, T_0$,
where $T_0=2\pi /\omega_0$ is the nanoresonator period.
One can see that as expected from the standard theory of stroboscopic QND
measurements, \cite{Braginsky2,Thorne}
there are sharp resonances at $\omega =2\,\omega_0/n$.
In the case of full modulation, $A_{mod}=1$, shown in Fig.\
\ref{fig-strob1}(a), the resonances have equal height; however,
their width decreases with $n$.
According to the QND idea, the squeezing should significantly decrease if
measurement is not switched completely off between the measurement pulses.
Comparing Figs.\ \ref{fig-strob1}(a) and \ref{fig-strob1}(b) we see
that the on/off ratio even as large as $10^3$ leads to a considerable
decrease of ${\cal S}$ [obviously, the effect of finite on/off ratio
becomes more important with decrease of $\delta t/T_0$].
Another consequence
of finite on/off ratio is the decrease of the resonant peak height
at $\omega =2\,\omega_0/n$ with $n$.

\begin{figure}
\centering
\includegraphics[width=2.7in]{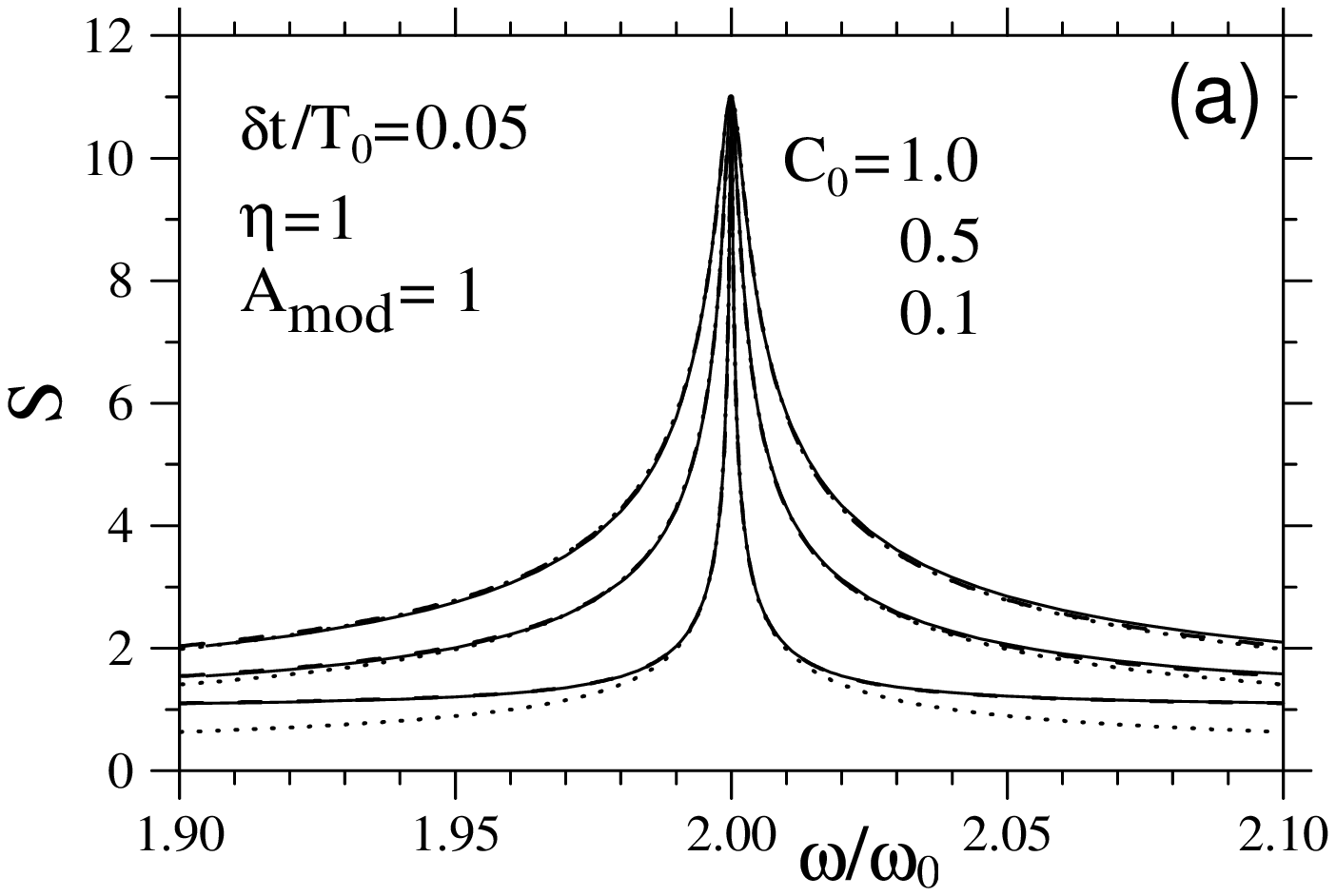}
\includegraphics[width=2.7in]{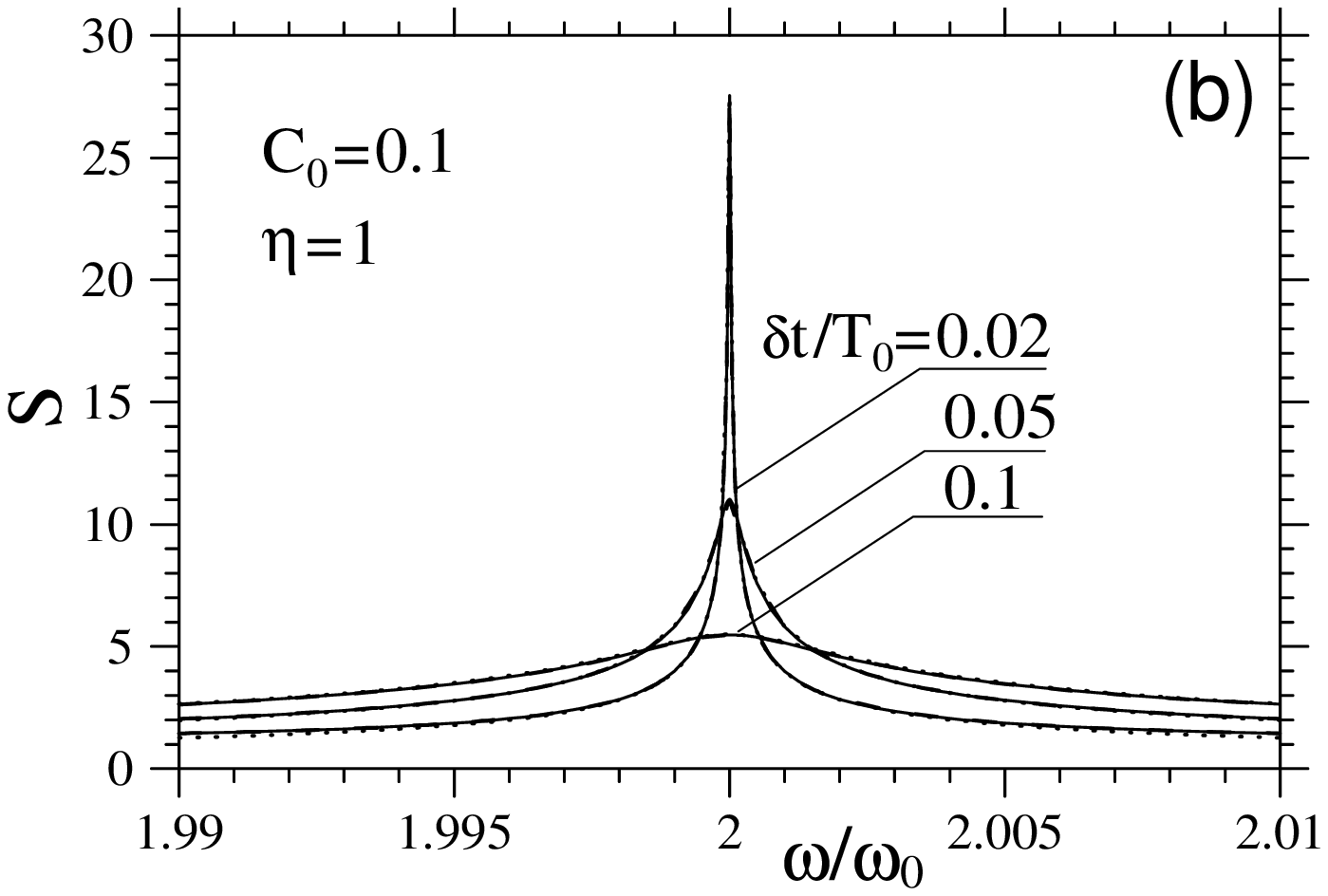}
\includegraphics[width=2.7in]{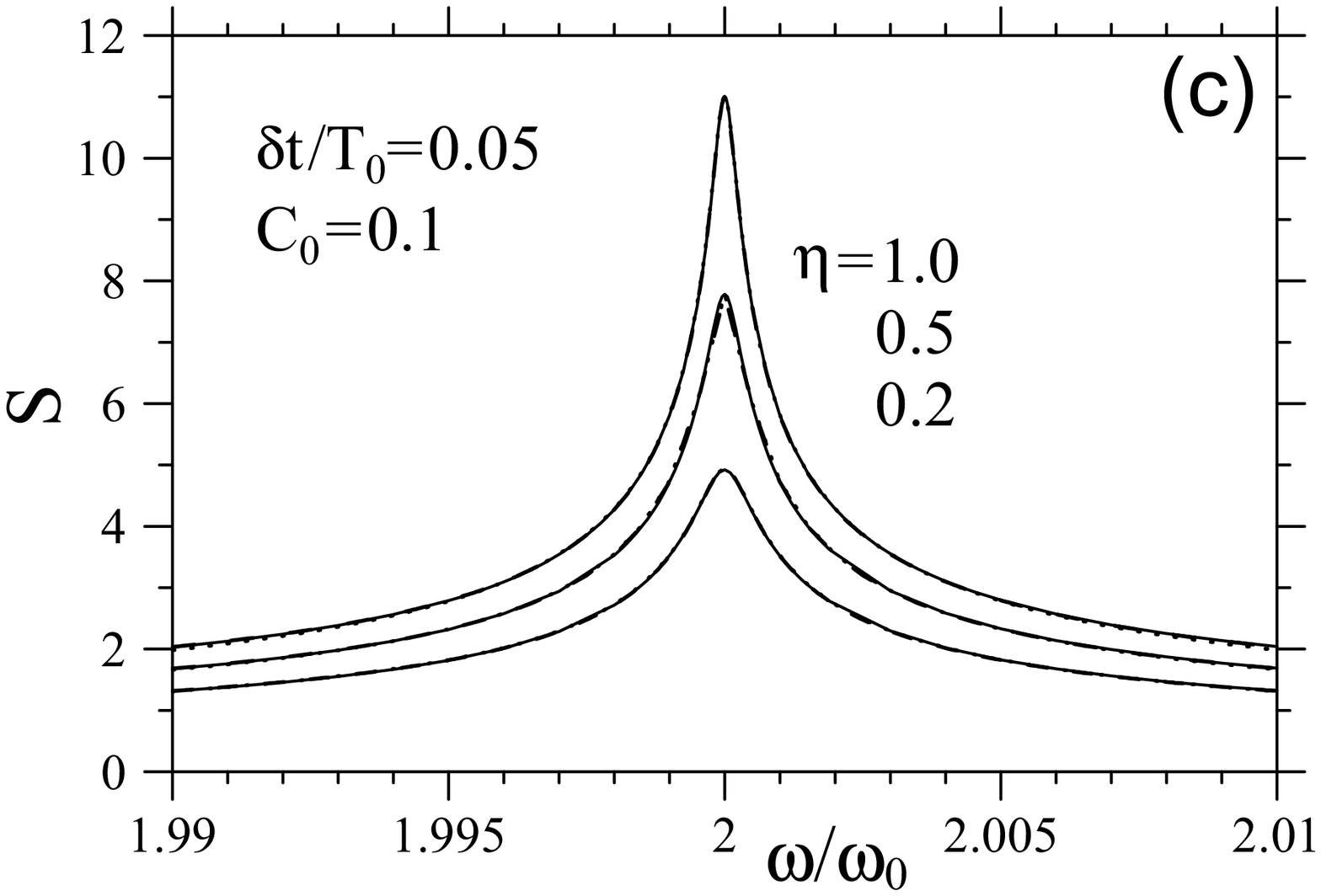}
% where an .eps filename suffix will be assumed under latex,
% and a .pdf suffix will be assumed for pdflatex
\caption{
The shape of the squeezing peak ${\cal S}(\omega )$ at
$\omega \simeq 2\,\omega_0$ for stroboscopic modulation, for several values
of (a) coupling with detector ${\cal C}_0$, (b) pulse duration $\delta t$
($T_0=2\pi/\omega_0$ is the resonator period), and (c) quantum efficiency
of measurement $\eta$.
Solid lines show numerical results, dashed lines (practically
indistinguishable
from the solid lines) are the analytical results given by Eqs.\
(\protect\ref{res-shape-str}) and (\protect\ref{S-A}), and the dotted
lines are calculated using the simplified equation (\protect\ref{Sq-A-str}).
The height of the squeezing peak is proportional to
$\sqrt{\eta}/\delta t$
[Eq.\ (\protect\ref{S-peak-str})] while its width is proportional to
${\cal C}_0(\delta t)^3/n^2\sqrt{\eta}$ [Eq.\ (\protect\ref{S-width-str})].
}
\label{fig-strob2}
\end{figure}

The results presented in Fig.\ \ref{fig-strob2}(a) show that for smaller
coupling ${\cal C}_0$ the peak height remains practically the same,
but the peak width decreases (this is the reason why we chose relatively
large coupling in Fig.\ \ref{fig-strob1} in order to have a noticeable peak
width). For smaller pulse duration $\delta t$, the squeezing peak
becomes higher and narrower [Fig.\ \ref{fig-strob2}(b)],
while the detector nonideality makes the peak lower and wider
[Fig.\ \ref{fig-strob2}(c)].
All these dependences will be confirmed by the analytical results
discussed below and shown in Fig.\ \ref{fig-strob2} by dashed and dotted
lines (dashed lines show more accurate results while dotted lines correspond
to a simpler formula).

\subsection{Analytical results for squeezing ${\cal S}$}

\subsubsection{Evolution of the state purity}

        Before discussing the analytical results for squeezing, let us
briefly discuss the evolution of the state purity,
$\mbox{Tr}(\rho^2) =(\hbar /2)/\sqrt{D_xD_p-D_{xp}^2}=1/\sqrt{u}$,
where $u=d_x d_p-d_{xp}^2$.
From Eqs.\ (\ref{eq-mDx-dles})--(\ref{eq-mDxp-dles})
(with $Q=\infty$ and $\gamma_{add}=0$)
 it is easy to derive the equation
        \begin{equation}
    \dot{u} = \omega_0 {\cal C}_0 |f(t)| d_x (\eta^{-1} -u)  .
        \label{dot-u}
        \end{equation}
Since ${\cal C}_0$ and $d_x$ are positive, the asymptotic solution
of this equation is obviously $u=1/\eta$ and therefore the state purity
reaches the asymptote $\mbox{Tr}(\rho^2)= \sqrt{\eta}$. In particular,
in the case of ideal detector, $\eta =1$, the state eventually becomes
pure (similar to the case of a qubit measurement \cite{Kor-99-01}).
As will be discussed later, the typical purification time is
comparable to the time of reaching the asymptotic regime for variances
$d_x$ and $d_p$.

\subsubsection{Analytics for harmonic modulation}

        For simplicity in this subsection we consider the harmonic modulation
(\ref{harm-mod}) of the measurement strength only with 100\% modulation,
$A_{mod}=1$ (which leads to maximum squeezing), and we still assume
$Q=\infty$.
        Without measurement, ${\cal C}_0 =0$,
Eqs.\ (\ref{eq-mDx-dles})--(\ref{eq-mDxp-dles})  have the solution
        \begin{eqnarray}
&& d_x(t)=\sqrt{\eta^{-1}+A^2} - A\,\cos(2\omega_0 t + \varphi) \, ,
\label{dx-str}\\
&& d_p(t)=\sqrt{\eta^{-1}+A^2} + A\,\cos(2\omega_0 t + \varphi) \, ,
\label{dp-str}\\
&& d_{xp}(t)=A\,\sin(2\omega_0 t + \varphi) \, ,
        \label{dxp-str}
        \end{eqnarray}
with arbitrary amplitude $A$ and phase $\varphi$. (Notice that these
equations satisfy the asymptotic condition $\mbox{Tr}\rho^2=\sqrt{\eta}$.)
For weak coupling,
${\cal C}_0/\eta \ll 1$, and harmonic modulation (\ref{harm-mod}) in the
vicinity of the resonance, $\omega \simeq 2\,\omega_0$,
it is natural to look for the asymptotic solution of
Eqs.\  (\ref{eq-mDx-dles})--(\ref{eq-mDxp-dles}) in the form
(\ref{dx-str})--(\ref{dxp-str}) with $2\omega_0$ replaced by $\omega$
(actually, $A$ and $\varphi$ vary in time with frequency $\omega$, but
variations are negligible at ${\cal C}_0/\eta \ll 1$).

        To find $A$ and $\varphi$, we substitute these equations into the
equation
$\int_{-\pi/\omega}^{\pi/\omega} f(t) (\eta^{-1} - d_x^2 - d_{xp}^2) \,
dt = 0$ which follows from the stationarity condition,
$\int_{-\pi/\omega}^{\pi/\omega} (\dot{d}_{x} + \dot{d}_{p})\, dt =0$,
and Eqs.\ (\ref{eq-mDx-dles})--(\ref{eq-mDp-dles}). This gives us
the relation
        \begin{equation}
A = \frac{1}{2}\,\sqrt{\eta^{-1} + A^2}\, \cos{\varphi} .
        \label{A-cosine}
        \end{equation}
We find numerically and analytically (see below)
that $\varphi =0$ at the resonance, $\omega = 2\,\omega_0$.
(This is quite natural: smaller $d_x$ correspond to
larger measurement strength.) Then from
Eq.\ (\ref{A-cosine}) we find $A=1/\sqrt{3\eta}$ and therefore
        \begin{equation}
{\cal S} (2\omega_0) = \sqrt{3\eta}
        \label{S-max-harm}
        \end{equation}
since the maximum squeezing ${\cal S}$ and the amplitude $A$ are related
as
        \begin{equation}
{\cal S}= \eta ( A+\sqrt{A^2+\eta^{-1}} ) .
        \label{S-A}
        \end{equation}
This result confirms the numerical result for the peak height in Fig.\ 1.

        To find the shape of the resonant peak, we need one more
equation relating $A$ and $\varphi$. It can be obtained by deriving equation
for $\ddot{d}_{xp} (t)$ from Eqs.\ (\ref{eq-mDx-dles})--(\ref{eq-mDxp-dles}),
and equating the $\sin (\omega t +\varphi )$ component for the two sides of
the
equation (assuming ${\cal C}_0/\eta \ll 1$ and $\omega \approx 2\,\omega_0$).
In this way we obtain
        \begin{equation}
(4 \omega_0^2 - \omega^2)\,A  =
 \eta^{-1} {\cal C}_0\,\omega_0^2 \sin{\varphi} \, .
         \label{2eqn-cos}
        \end{equation}
In particular, this proves that $\varphi =0$ at $\omega =2\,\omega_0$.
Combining Eqs.\ (\ref{A-cosine}) and (\ref{2eqn-cos}) we find the amplitude
$A$ as
        \begin{equation}
A(\omega) = \sqrt{ \frac{2/\eta }{3+g(\omega) +
\sqrt{g^2(\omega) + 10 g(\omega) +9}} } \, ,
        \label{Sq-A-cos}
        \end{equation}
where $ g(\omega) = 16\eta (2-\omega/\omega_0)^2/{\cal C}_0^2$.
The corresponding analytical result for squeezing ${\cal S}$ is obtained
via Eq.\ (\ref{S-A}).
This result is shown by the dashes lines
in Figs.\ \ref{fig-cos}(a) and \ref{fig-cos}(b), which practically coincide
with the solid lines representing the numerical results.
Notice that the linewidth of the peak is proportional to
${\cal C}_0/\sqrt{\eta}$;
 away from the resonance $A$ decreases to zero, and ${\cal S}$
approaches ${\cal S}=\sqrt{\eta}$, which is the same as for the case
without modulation. \cite{DohertyJacobs} The analytical result for
${\cal S}(\omega )$ works well for coupling ${\cal C}_0$ up to approximately
0.3; for larger ${\cal C}_0$ there is a noticeable difference from
the numerical result as seen in Fig.\ \ref{fig-cos}(a).
        It is curious that rather complex shape of the resonance peak
given by Eqs.\ (\ref{S-A}) and (\ref{Sq-A-cos}) is quite close to the square
root of the Lorentzian shape:
        \begin{equation}
{\cal S} (\omega ) \approx \sqrt{\eta } \left( 1 +
\frac{\sqrt{3}-1}
{ \sqrt{1 + 3 [ (\omega-2 \omega_0)/\Delta\omega ]^2}} \right)
        \label{sqrt-Lorenz}
        \end{equation}
with half-width at half-height
$\Delta\omega \simeq 0.63\, \omega_0\, {\cal C}_0/\sqrt{\eta}$.

\subsubsection{Analytics for stroboscopic modulation}

        In the case of stroboscopic modulation (\ref{strob-mod}) of the
measurement strength (in this subsection we assume full modulation,
$A_{mod}=1$, and still neglect $Q$-factor),
the variances $d_x$, $d_p$ and $d_{xp}$ should
follow Eqs.\ (\ref{dx-str})--(\ref{dxp-str}) during the ``off'' phase of
the modulation, while during the measurement pulse of duration
$\delta t$ the parameters $A$ and $\varphi$
slowly change (we again assume the weak coupling limit) in accordance with
Eqs.\ (\ref{eq-mDx-dles})--(\ref{eq-mDxp-dles}). In particular,
close to the $n$th resonant peak of Fig.\ \ref{fig-strob1}(a),
$\omega \approx 2\,\omega_0/n$,
the phase $\varphi$ should change during the pulse by the small amount
        \begin{equation}
\delta \varphi = -2\omega_0\,(2\pi /\omega) + 2\pi\,n \approx
 \pi\, n^2\, (\omega/\omega_0 - 2/n)
        \label{deltphi}
        \end{equation}
 in order to match $2\pi/\omega$ periodicity of the asymptotic
solution with the periodicity of free oscillations
(\ref{dx-str})--(\ref{dxp-str}). On the other hand, $\delta \varphi$ can
be found from the equation
        \begin{equation}
\dot{\varphi} =-4\omega_0 {\cal C}_0 \eta^{-1} |f(t)| d_{xp}/
        [(d_p-d_x)^2+4d_{xp}^2]
        \label{dot-phi}
        \end{equation}
which follows from from Eqs.\ (\ref{eq-mDx-dles})--(\ref{eq-mDxp-dles}).
Integrating Eq.\ (\ref{dot-phi}) within the pulse interval
$|t|\leq \delta t/2$ using Eqs.\ (\ref{dx-str})--(\ref{dxp-str}) in which
 $A$ and $\varphi$ are assumed constant, we obtain
$\delta \varphi = - {\cal C}_0 \sin (\omega_0\delta t)/\eta A$.
Combining this result with Eq.\ (\ref{deltphi}) we obtain an equation
relating $A$ and $\varphi$:
        \begin{equation}
\pi n^2 A (\omega/\omega_0 - 2/n) = \eta^{-1}{\cal C}_0
\sin (\omega_0 \delta t)\, \sin \varphi \, .
        \label{2eqn}
        \end{equation}

To obtain one more equation for $A$ and $\varphi$, we use the condition
$\int_{-\delta t/2}^{\delta t/2} (\dot{d}_x+\dot{d}_p) \, dt =0$.
Expressing the derivative $\dot{d}_x+\dot{d}_p$ from Eqs.\
(\ref{eq-mDx-dles})--(\ref{eq-mDp-dles}) and using Eqs.\
(\ref{dx-str})--(\ref{dxp-str}), we get the equation
        \begin{equation}
A\, \omega_0\,\delta t = \sqrt{\eta^{-1}+A^2}\ \sin(\omega_0 \delta t)\,
\cos{\varphi} \, .
        \label{1eqn}
        \end{equation}
Equations (\ref{2eqn}) and (\ref{1eqn}) are sufficient to find $A$
for the $n$th resonance, though the expression is quite long:
        \begin{equation}
A^2(\omega) = \frac{2\, \eta^{-1}\, \sin^2 (\omega_0 \delta t)}
{B(\omega) + \sqrt{B^2(\omega) + 4\, \tilde{g}(\omega) \sin^2
(\omega_0 \delta t)} }
\label{res-shape-str} ,
\end{equation}
where $B(\omega)=\tilde{g}(\omega ) + (\omega_0 \delta t)^2 - \sin^2
(\omega_0 \delta t)$
and $\tilde{g}(\omega )=\pi^2 n^2 (2/n-\omega/\omega_0)^2 \eta/{\cal C}_0^2$.
The squeezing ${\cal S}$ is obtained from this result using Eq.\ (\ref{S-A}).
The corresponding analytical curves are plotted in Fig.\
\ref{fig-strob2} by the
dashed lines which practically coincide with the numerical results shown
by the solid lines. One can see that the analytics works well even
for ${\cal C}_0=1$, even though we assumed ${\cal C}_0\ll 1$ for
the derivation.

        The value of squeezing at $\omega =2\,\omega_0/n$ (peak height) can
be obtained from Eq.\ (\ref{res-shape-str}), but it is easier to use
Eq.\ (\ref{1eqn}) with $\varphi =0$ [which follows from Eq.\ (\ref{2eqn})],
that leads to the result
        \begin{equation}
{\cal S}(2 \omega_0/n) = \sqrt{\eta}\,
\sqrt{\frac{\omega_0 \delta t + \sin (\omega_0 \delta t)}
{\omega_0 \delta t - \sin (\omega_0 \delta t)}} \, .
        \label{Sq-max-strob}
        \end{equation}

        The analytical results simplify in the case of
short pulses, $\delta t \ll T_0=2\pi /\omega_0$, then
        \begin{equation}
A^2(\omega) =
   \frac{6/(\omega_0 \delta t )^2\eta}
{1+\sqrt{1+\left[\frac{6\pi\sqrt{\eta}\, n^2 (\omega-2\omega_0/n) }
{{\cal C}_0 (\delta t)^3 \omega_0^4} \right]^2  }} \, ,
        \label{Sq-A-str}
        \end{equation}
which corresponds to the peak squeezing
        \begin{equation}
{\cal S} (2\omega_0/n) = 2\sqrt{3\eta}/\omega_0\delta t
        \label{S-peak-str}
        \end{equation}
(since ${\cal S}=2\eta A$ for ${\cal S} \gg 1$, and
$A=\sqrt{3}/\omega_0\delta t\sqrt{\eta}$),
while the half-width at half-height of ${\cal S} (\omega )$  is
        \begin{equation}
\Delta \omega = 2{\cal C}_0 (\delta t)^3\omega_0^4/\pi n^2\sqrt{3\eta}\, .
        \label{S-width-str}
        \end{equation}
The curves calculated using Eq.\ (\ref{Sq-A-str}) are shown in Fig.\
\ref{fig-strob2}
by the dotted lines. There is a noticeable difference
from the numerical results away from the resonance; however, the main
part of the peak is fitted quite well.

    Notice that in the case of exact resonance, $\omega =2\,\omega_0/n$,
the smallest $x$-width of the wavepacket is achieved at the middle
of the measurement pulse, and at this point $d_x=1/{\cal S}$. However,
$d_x$ increases considerably even within the duration of the pulse,
so that the maximum value $d_{x,max}^{\delta t} = 4/{\cal S}$
within the pulse is at its onset and end, while $d_x$ averaged over
the pulse duration is $\overline{d_x^{\delta t}}=2/{\cal S}$.

        \subsection{Timescale of squeezing buildup}

An important question is how fast the squeezing approaches
its asymptotic value calculated in Subsections A and B.
In this Subsection we analyze the duration of the transient period
of squeezing buildup (see Fig.\ \ref{fig-time}) for stroboscopic
modulation $f(t)$ with $A_{mod}=1$ and $\delta t/T_0 \ll 1$
at resonance, $\omega =2\,\omega_0/n$, assuming $Q=\infty$.

        Let us start with the standard QND case of instantaneous
imprecise measurements, \cite{Braginsky2,Thorne} which corresponds
to the formal limit $\delta t \rightarrow 0$, ${\cal C}_0\rightarrow \infty$,
while ${\cal C}_0\delta t =\mbox{const}$. Each measurement changes
the resonator density matrix by multiplying it by a Gaussian function
[see Eqs.\ (\ref{Bayes-gen}) and (\ref{I-Gauss})] with $x$-variance
$D=(\Delta x_{0})^2/{\cal C}_0 \omega_0 \delta t$. Since the free
resonator evolution in between the measurements can be neglected if the
measurements are separated by integer number of half-periods, the total
strength of repeated measurements adds up (product of two Gaussians
is a Gaussian with added inverse variances). Therefore for a Gaussian
initial state (\ref{Gaussian-den-mat}) the squeezing magnitude ${\cal S}_N$
after $N$ measurements is
        \begin{equation}
{\cal S}_{N} = N\,{\cal C}_0 \omega_0 \delta t + {\cal S}_{0} .
        \label{Sq-strong}
        \end{equation}

\begin{figure}
\centering
\includegraphics[width=2.8in]{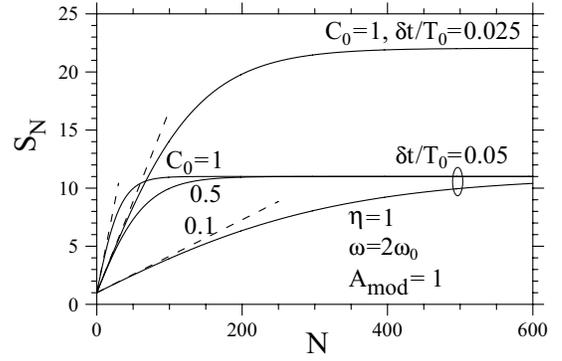}
% where an .eps filename suffix will be assumed under latex,
% and a .pdf suffix will be assumed for pdflatex
\caption{
Gradual buildup of squeezing ${\cal S}$ with number of measurement
pulses $N$ starting from the ground state for several values of
coupling ${\cal C}_0$ and pulse duration $\delta t$. Solid lines are the
numerical results, dashed lines correspond to Eq.\ (\ref{Sq-strong}).
}
\label{S(N)}
\end{figure}

        While for instantaneous measurements the magnitude of squeezing
accumulates indefinitely, for a continuous stroboscopic measurement with
finite $\delta t$ the quantum back-action cannot be avoided completely,
so the squeezing increases as Eq.\ (\ref{Sq-strong})
only during the initial transient period (Fig.\ \ref{S(N)}) and
then saturates at the asymptotic level analyzed in Subsections A and B.
Comparing the asymptotic value ${\cal S}_\infty$ given by
Eq.\ (\ref{S-peak-str})
with the increase rate from Eq.\ (\ref{Sq-strong}) and
neglecting initial value ${\cal S}_0$, we obtain the estimate
        \begin{equation}
   N_b = \frac{2\sqrt{3\eta}}{{\cal C}_0(\omega_0\delta t)^2}
        \label{N_b}
        \end{equation}
for the number of measurement pulses necessary for almost complete
buildup of squeezing
(of course the numerical factor $2\sqrt{3}$ is not really important here).

        Since the saturation of ${\cal S}_N$ is a gradual process,
let us also analyze analytically the $N$-dependence for $N\agt N_b$
when ${\cal S}_N$ is already close to ${\cal S}_\infty$. Let us start with
Eqs.\ (\ref{dx-str})--(\ref{dxp-str}) assuming that the asymptotic
purity $\mbox{Tr}(\rho^2)=\sqrt{\eta}$ is already reached but the
parameter $A$ still changes with $N$.
Following the derivation used in Subsection B.3, we combine Eqs.\
(\ref{dx-str})--(\ref{dp-str}) with
(\ref{eq-mDx-dles})--(\ref{eq-mDp-dles}) and obtain
$ \dot{d_x} + \dot{d_p} = 2 A \dot{A} /\sqrt{\eta^{-1} + A^2}
= \omega_0\, {\cal C}_0 |f(t)| (\eta^{-1} - d_x^2 - d_{xp}^2 )$.
Integrating this equation over the measurement pulse duration
(assuming ${\cal C}_0\ll 1$ and $\varphi =0$) we find the corresponding
small change of the parameter $A$:
     \[
\frac{\Delta A}{\Delta N} = {\cal C}_0\sqrt{\eta^{-1} + A^2}\,
[\sqrt{\eta^{-1}+A^2}\,\sin{\omega_0\delta t}-A \omega_0\delta t] .
     \]
Translating this equation into evolution of squeezing and assuming
$|{\cal S}_N-{\cal S}_\infty| \ll{\cal S}_\infty$, $\omega_0\delta t \ll 1$,
$A^2 \gg \eta^{-1}$, we obtain
        \begin{equation}
\frac{\Delta {\cal S}_N}{\Delta N} =
-\frac{{\cal C}_0\,(\omega_0 \delta t)^2}{\sqrt{3\eta}}\,
({\cal S}_N -{\cal S}_\infty),
        \label{deltaS-change}
        \end{equation}
which shows the exponential approach of the squeezing ${\cal S}_N$
towards ${\cal S}_\infty$ as $\exp (-2N/N_b)$, i.e. the typical
number of measurements necessary to reach the asymptotic value
is similar to what was found from initial part of the transient,
Eq.\ (\ref{N_b}).

        It is interesting to notice that the timescale of the
purity factor saturation (see Fig.\ \ref{fig-time})
is similar to the timescale of squeezing saturation.
Using Eq.\ (\ref{dot-u}) for $u \equiv d_x d_p -d_{xp}^2 =
1/ \mbox{Tr}^2(\rho^2)$
and approximating average $d_x$ within the pulse duration as
$\overline{d_x^{\delta t}} = [1/\eta A + A (\omega_0 \delta t)^2/4]/2$,
which becomes
$\overline{d_x^{\delta t}} \simeq 2/{\cal S}_\infty$ close to saturation,
we obtain
        \begin{equation}
\frac{\Delta u}{\Delta N} = -\frac{{\cal C}_0\,(\omega_0 \delta t)^2}
{\sqrt{3\eta}}\, (u - 1/\eta ) .
        \label{deltaU-change}
        \end{equation}
Therefore, similar to ${\cal S}_N$ behavior, $u$ also approaches asymptote
as $\exp (-2N/N_b)$. Far from saturation we expect
$\overline{d_x^{\delta t}} > 2/{\cal S}_\infty$, and therefore a larger
initial rate of reaching the asymptote.

\begin{figure}
\centering
\includegraphics[width=2.8in]{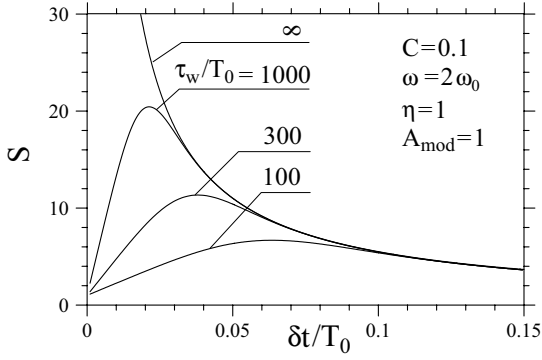}
% where an .eps filename suffix will be assumed under latex,
% and a .pdf suffix will be assumed for pdflatex
\caption{
Squeezing ${\cal S}$ as a function of the pulse duration $\delta t$
for stroboscopic measurements with a particular ``waiting time''
$\tau_w$ allowed for squeezing buildup.
}
\label{tau_w}
\end{figure}

        Finite time scale of squeezing buildup is important if
an allowed experimental ``waiting time'' $\tau_w$ is limited.
Figure \ref{tau_w} shows the squeezing ${\cal S}$ as a function of
measurement pulse duration $\delta t$ for several values of
$\tau_w$ (initial state is chosen to be the ground state).
While the upper line ($\tau_w=\infty$) corresponds to Eq.\
(\ref{S-peak-str}) and increases indefinitely at small $\delta t$,
the squeezing for finite waiting time $\tau_w$ reaches maximum
at an optimum pulse duration $\delta t$. For smaller $\delta t$
the squeezing buildup is too slow [see Eq.\ (\ref{N_b})] and
the squeezing is limited by the accumulated measurement strength
${\cal C}_0 (2\tau_w/T_0) (\delta t/T_0)$, while for larger
$\delta t$ the limiting factor is too strong back-action.

Numerical calculations show that this maximum squeezing is fitted well
by the formula
        \begin{equation}
{\cal S}_{max} \approx 2.0 \sqrt{{\cal C}_0 \tau_{w}/T_0 }
\hspace{0.5cm}  (\omega =2\, \omega_0, \eta =1).
        \label{S-tau_w1}
        \end{equation}
[For example, this fitting formula underestimates the numerical results
for ${\cal C}_0=0.1$ (${\cal C}_0=0.5$) by
$9\%$ ($4.6\%$) for $\tau_{wait}/T_0=30$ and by $2.2\%$ ($1.2\%$)
for $\tau_{wait}/T_0=1000$.]

        For an analytical estimate of ${\cal S}_{max}$ let us assume that
optimum $\delta t$ corresponds to the condition of squeezing buildup time
being comparable to the waiting time, $2\tau_w/T_0 \simeq N_b$ [see
Eq.\ (\ref{N_b})]. Then the optimum is achieved at
        \begin{equation}
\delta t_{opt}/T_0 \simeq 0.21 \eta^{1/4}/({\cal C}_0\tau_w/T_0)^{1/2}
        \label{dt-opt}
        \end{equation}
(which is well confirmed by results in Fig.\ \ref{tau_w}) and the
corresponding ${\cal S}_{max}$ calculated from Eq.\ (\ref{S-peak-str}) is
        \begin{equation}
  {\cal S}_{max} \simeq 2 (3\eta)^{1/4} \sqrt{{\cal C}_0\tau_w/T_0},
        \label{S-tau_w2}
        \end{equation}
which differs from the numerical result (\ref{S-tau_w1}) only by
a factor $\approx 1.3$.

        It is tempting to guess that the effect of finite $Q$-factor
(at least for zero temperature of environment) can be described by
a similar formula with $\tau_w$ replaced by $QT_0$
(so that ${\cal S}_{max} \simeq \eta^{1/4} \sqrt{{\cal C}_0 Q}$)
 since $\tau_w$
is naturally restricted by the resonator damping time. However,
as will be seen in the next Subsection, this gives only
an upper bound and finite $Q$-factor actually leads to a significantly
smaller value of ${\cal S}_{max}$.

        \subsection{Effects of finite $Q$-factor and environment
temperature}

        In this Subsection we analyze effects of finite quality factor
$Q$ of the nanoresonator and environment temperature $T$ for stroboscopic
measurement with $\omega =2\,\omega_0/n$ and $A_{mod}=1$. (Extra dephasing
$\gamma_{add}$ is equivalent to increase of $T$.)

        Numerical solution of Eqs.\ (\ref{eq-mDx-dles})--(\ref{eq-mDxp-dles})
with a finite $Q$-factor ($Q\gg 1$) shows that as expected the squeezing
${\cal S}$ decreases at sufficiently small $Q$, and
higher temperature  also decreases ${\cal S}$.
While for $Q=\infty$ the squeezing does not depend on coupling
with detector ${\cal C}_0\lesssim 1$ for a fixed pulse duration $\delta t$
 [see Fig.\ \ref{fig-strob2}(a) and
Eq.\ (\ref{S-peak-str})], for a finite $Q$ the squeezing starts to decrease
for too small ${\cal C}_0$, since coupling with detector competes with
coupling to environment.
 [The effect is to some
extent similar to the effect of $A_{mod}<1$; in particular, the squeezing
at $\omega =2\,\omega_0/n$ decreases stronger with $n$ as in Fig.\
\ref{fig-strob1}(b).] Notice that for infinite $Q$ the environment
temperature is not important since nanoresonator is not coupled to
the environment and the evolution is determined by coupling with detector
only.

     For an analytical analysis let us mention first that the asymptotic
purity $\mbox{Tr}(\rho^2)=1/\sqrt{u}$ is no longer equal to $\sqrt{\eta}$,
since Eq.\ (\ref{dot-u}) should be replaced by
        \begin{equation}
\frac{\dot{u}}{\omega_0} =  {\cal C}_0 |f(t)| d_x (\eta^{-1} -u)
- \frac{2u}{Q} + \frac{2d_x}{Q}
\coth\frac{\hbar\omega_0}{2 T} .
        \label{purity-Q}
        \end{equation}
For $Q\gg 1$ we can neglect small asymptotic oscillations of $u$ and
assume a practically constant asymptotic value $\tilde u$.
Since the average of Eq.\ ({\ref{purity-Q}}) over the oscillation period
should be equal to zero in the asymptotic regime, we can find $\tilde u$
from equation
        \begin{equation}
{\cal C}_0 \overline{d_x^{\delta t}}\, \frac{\omega \delta t}{2\pi} \,
(\eta^{-1} -{\tilde u})
- \frac{2}{Q}  \left( \tilde{u}
- \overline{d_x} \coth \frac{\hbar\omega_0}{2 T} \right) =0 ,
        \end{equation}
where $\overline{d_x^{\delta t}}$ is $d_x$ averaged over the pulse duration
while $\overline{d_x}$ is averaged over the whole period.

        To find $\overline{d_x^{\delta t}}$ and $\overline{d_x}$ we
use Eqs.\ (\ref{dx-str})--(\ref{dxp-str}) which are still applicable for
the asymptotic oscillations of $d_x$, $d_p$, and $d_{xp}$ if
$\eta^{-1}$ in these equations is replaced with $\tilde u$.
Still assuming no phase shift $\varphi$ in case of exact resonance
$\omega =2\,\omega_0/n$ (this has been confirmed numerically), we
obtain
        \begin{eqnarray}
&& \hspace{-0.4cm}
 (\eta^{-1}-{\tilde u})\, [ \sqrt{{\tilde u}+A^2}\, \omega_0\delta t
 -A \sin (\omega_0\delta t) ]
\nonumber\\
&& \hspace{-0.4cm}
+ (2\pi n /{\cal C}_0 Q)\, [\coth (\hbar\omega_0/2T)
\sqrt{{\tilde u}+A^2} -{\tilde u} ] = 0 \, .
        \label{1st-eq-uA}
        \end{eqnarray}
One more equation which relates ${\tilde u}$ and $A$ follows from
zero average of $\dot{d_x}+\dot{d_p}$ in the stationary regime.
Using Eqs.\ (\ref{eq-mDx-dles})--(\ref{eq-mDxp-dles}) and
modified Eqs.\ (\ref{dx-str})--(\ref{dxp-str}) we find
        \begin{eqnarray}
&&  (\eta^{-1}-2 A^2-{\tilde u})\,\omega_0\delta t
+ 2 A\sqrt{{\tilde u}+A^2}\,\sin (\omega_0\delta t)
\nonumber\\
&&
- (2\pi n/{\cal C}_0 Q)\,[ \sqrt{{\tilde u}+A^2}-
\coth (\hbar\omega_0/2T) ] = 0 \, .
        \label{2nd-eq-uA}
        \end{eqnarray}

        We have checked that the squeezing
${\cal S}= ( A+\sqrt{A^2+{\tilde u}})/{\tilde u}$
[see Eq.\ (\ref{S-A})] calculated from the numerical solution of
Eqs.\ (\ref{1st-eq-uA}) and (\ref{2nd-eq-uA}) practically coincides
with results from direct solution of Eqs.\
(\ref{eq-mDx-dles})--(\ref{eq-mDxp-dles}) for ${\cal C}_0 Q\agt 10$.
It is also easy to check that in the limit $Q=\infty$ Eq.\
(\ref{2nd-eq-uA}) transforms into Eq.\ (\ref{1eqn}); therefore
we reproduce our previous results (\ref{Sq-max-strob}) and
(\ref{S-peak-str}) for squeezing.

\begin{figure}
\centering
\includegraphics[width=2.8in]{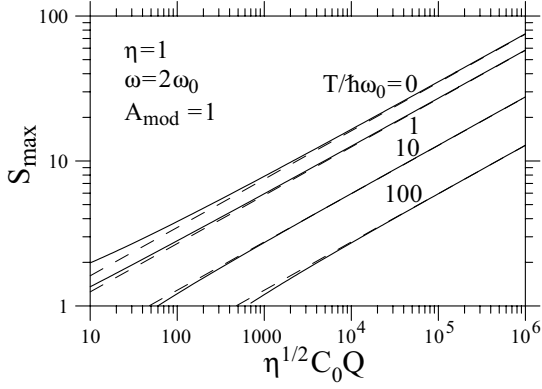}
% where an .eps filename suffix will be assumed under latex,
% and a .pdf suffix will be assumed for pdflatex
\caption{
The squeezing ${\cal S}$ maximized over the stroboscopic pulse duration
$\delta t$ as a function of nanoresonator $Q$-factor (multiplied by
coupling with detector ${\cal C}_0$ and square root of efficiency $\eta$)
for several values of nanoresonator
temperature $T$. Solid lines are the numerical results, dashed lines
correspond to the fitting formula (\protect\ref{Q-fit}).
}
\label{fig-Q}
\end{figure}

        Solid lines in Fig.\ \ref{fig-Q} show the dependence of maximum
squeezing (optimized over the pulse duration $\delta t$) as a function
of the product ${\cal C}_0 Q$ for several temperatures of the environment,
calculated numerically using Eqs.\ (\ref{1st-eq-uA}) and (\ref{2nd-eq-uA})
for $\omega =2\omega_0$.
These results are fitted well by the formula
        \begin{equation}
{\cal S}_{max} = \frac{3}{4} \,  \eta^{1/6} \left[
\frac{{\cal C}_0 Q}{\coth (\hbar\omega_0/2T)} \right]^{1/3} ,
        \label{Q-fit}
        \end{equation}
shown by dashed lines in Fig.\ \ref{fig-Q}.
As we see the scaling $({\cal C}_0 Q)^{1/3}$ is more restrictive
than scaling $({\cal C}_0 Q)^{1/2}$ which could be guessed from Eq.\
(\ref{S-tau_w2}).

    For an analytical estimate of ${\cal S}_{max}$ let us start with
high-temperature case, $T\gg \hbar\omega_0$. The thermal noise contributes
to the increase of $x$-variance (due to random walk) crudely as
\cite{BraginskyKhalili} $\dot{d}_x= (2T/\hbar\omega_0)(\omega_0/Q)$.
In stationary state this increase is compensated by the squeezing buildup
contribution which can be estimated from Eq.\ (\ref{Sq-strong}) as
$\dot{d}_x=-(2/nT_0){\cal C}_0\omega_0\delta t/{\cal S}^2$,
assuming that ${\cal S}$ is mainly limited
by the effect of $Q$-factor [so ${\cal S}$ is much smaller than the value
given by Eq.\ (\ref{S-peak-str}) for infinite $Q$].
Equating two contributions, we obtain
${\cal S}^2=(2/n)(\delta t/T_0)\, {\cal C}_0 Q (\hbar\omega_0 /2T)$.
Addition of quantum noise to the thermal noise leads to replacement of
$2T/\hbar\omega_0$ by $\mbox{coth}(\hbar\omega_0/2T)$, which gives
    \begin{equation}
{\cal S} = \left[ \frac{2}{n}\, \frac{\omega_0\delta t}{2\pi} \,
\frac{{\cal C}_0 Q}{ \mbox{coth} (\hbar\omega_0 /2T)} \right]^{1/2} \, .
    \label{S-Q-an}
    \end{equation}
We have checked that numerical solutions of Eqs.\ (\ref{1st-eq-uA})
and (\ref{2nd-eq-uA}) practically coincide with this formula when
squeezing ${\cal S} \gg 1$ is mainly limited by $Q$-factor.
Finally, comparing this formula with the limitation for infinite $Q$
[Eq.\ (\ref{S-peak-str})] and optimizing over $\delta t$, we obtain
the estimate
${\cal S}_{max} = a \eta^{1/6} [{\cal C}_0 Q/n\,\mbox{coth}
(\hbar\omega_0/2T)]^{1/3}$ with the numerical factor $a \simeq 1.03$;
this factor is obviously supposed to overestimate the result of numerical
optimization, Eq.\ (\ref{Q-fit}).

        As follows from Eqs.\ (\ref{Q-fit}), (\ref{S-Q-an}),
and (\ref{S-peak-str}),
the effect of finite $Q$-factor is not important only when
both ${\cal C}_0 Q$ and ${\cal C}_0 Q \hbar \omega_0/T$ are much larger
than ${\cal S}^3/\sqrt{\eta} \sim \eta/(\omega_0\delta t)^3$.

\section{Quantum feedback of the packet center}

As shown in the previous Section, the $x$-width of the monitored Gaussian
wavepacket can be squeezed below the ground state width by applying periodic
modulation $|f(t)|$ of the measurement strength.
However, because of the measurement back-action, the center of
the wavepacket undergoes random evolution described by Eqs.\
(\ref{eq-m-x})--(\ref{eq-m-p}), and without feedback diffuses far away
from the origin. The diffusion is eventually limited either by damping
due to finite $Q$-factor or by very large (formally infinite in our
model) effective temperature (voltage) of the detector.
\cite{DohertyJacobs,MozyrskyMartin,Hopkins}
Even though the evolution of the wavepacket center can be monitored
using Eqs.\ (\ref{eq-m-x})--(\ref{eq-m-p}) in each realization of the process
and therefore the produced squeezed state is in principle useful for
applications, large fluctuations of the center position would clearly
lead to technical difficulties. The goal of this Section is to show
that the wavepacket center can be kept very close to origin
all the time using quantum feedback.

        The feedback is described by the force ${\cal F}$ in Eq.\
(\ref{eq-m-p}). Similar to Refs.\ \onlinecite{DohertyJacobs} and
\onlinecite{Hopkins} we choose the linear feedback of the form
        \begin{equation}
{\cal F} =-m\omega_0 \gamma_x \langle x\rangle - \gamma_p \langle p\rangle ,
        \label{fb-force}
        \end{equation}
where $\langle x\rangle$ and $\langle p\rangle$ are the continuously
monitored values. To analyze the feedback performance we characterize
\cite{DohertyJacobs,Hopkins}
the distribution of the packet center position $\langle x \rangle $
and center momentum $\langle p\rangle$ by the ensemble averages
(over realizations)
$\langle\langle x \rangle\rangle$ and $\langle\langle p \rangle\rangle$
and the variances $D_{\langle x\rangle}=\langle \langle x\rangle^2\rangle
-\langle\langle x\rangle\rangle^2$,
$D_{\langle p\rangle}=\langle \langle p\rangle^2\rangle
-\langle\langle p\rangle\rangle^2$,
$D_{\langle x\rangle\langle p\rangle}=\langle \langle x\rangle
\langle p\rangle \rangle -\langle\langle x\rangle\rangle
\langle\langle p\rangle\rangle $.
In the notation of doubled angle brackets the inner brackets
mean averaging with the density matrix $\rho$ in an individual realization
of the process,
while the outer brackets is averaging over realizations.
Notice that a natural characteristic of the total $x$-deviation of
the state from the origin is the sum $D_x+D_{\langle x\rangle} +
\langle\langle x\rangle\rangle^2$, so the feedback goal is
to ensure $D_{\langle x\rangle} + \langle\langle x\rangle\rangle^2
\lesssim D_x =(\Delta x_0)^2/{\cal S}$ to keep the squeezed state
sufficiently well centered.

        The equations for $\langle\langle \dot{x}\rangle\rangle$
and $\langle\langle \dot{p}\rangle\rangle$ derived from Eqs.\
(\ref{eq-m-x})--(\ref{eq-m-p}) lead to the
ensemble-averaged evolution
        \begin{equation}
\langle\langle \ddot{x} \rangle\rangle +
(\gamma_p+\omega_0/Q ) \langle\langle \dot{x} \rangle\rangle
+(\omega_0^2+\gamma_x\omega_0)\langle\langle x \rangle\rangle =0,
        \end{equation}
which shows that $\langle\langle x\rangle\rangle$ eventually relaxes
to zero for positive $\gamma_p$ even if $Q$-factor is infinite.

        Introducing dimensionless variances
$d_{\langle x\rangle}\equiv D_{\langle x \rangle} 2m \omega_0
/ \linebreak[0] \hbar$,
$d_{\langle p\rangle}\equiv  D_{\langle p\rangle} 2 /\hbar m \omega_0$, and
$d_{\langle x \rangle\langle p\rangle} \equiv
D_{\langle x\rangle\langle p\rangle} 2/\hbar $,
we derive \cite{DohertyJacobs,Hopkins}
 the following equations from Eqs.\ (\ref{eq-m-x})--(\ref{eq-m-p}):
        \begin{eqnarray}
&& \dot{d}_{\langle x\rangle}/ \omega_0 =
2 d_{\langle x\rangle\langle p\rangle} + {\cal C}_0 |f(t)| \, d_x^2 \, ,
\label{eq-cmDx-dles}\\
&& \dot{d}_{\langle p\rangle}/\omega_0 =
-2 d_{\langle x\rangle\langle p\rangle}
- 2 \mu F d_{\langle x\rangle\langle p\rangle} - 2 F d_{\langle p\rangle}
        \nonumber \\
&&\hspace{1.5cm}  + {\cal C}_0 |f(t)|\, d_{xp}^2
-(2/Q)\, d_{\langle p\rangle} \, ,
\label{eq-cmDp-dles}\\
&& \dot{d}_{\langle x\rangle\langle p\rangle}/\omega_0 =
 d_{\langle p\rangle} - d_{\langle x\rangle}
- \mu F d_{\langle x\rangle} - F d_{\langle x\rangle\langle p\rangle}
\nonumber\\
&&\hspace{1.5cm}
+ {\cal C}_0  |f(t)|\, d_x d_{xp}
-(1/Q) \, d_{\langle x\rangle\langle p\rangle} \, ,
        \label{eq-cmDxp-dles}
        \end{eqnarray}
where $F= \gamma_p/\omega_0$ and $\mu = \gamma_x/\gamma_p$
are the dimensionless feedback parameters.

\begin{figure}
\centering
\includegraphics[width=2.7in]{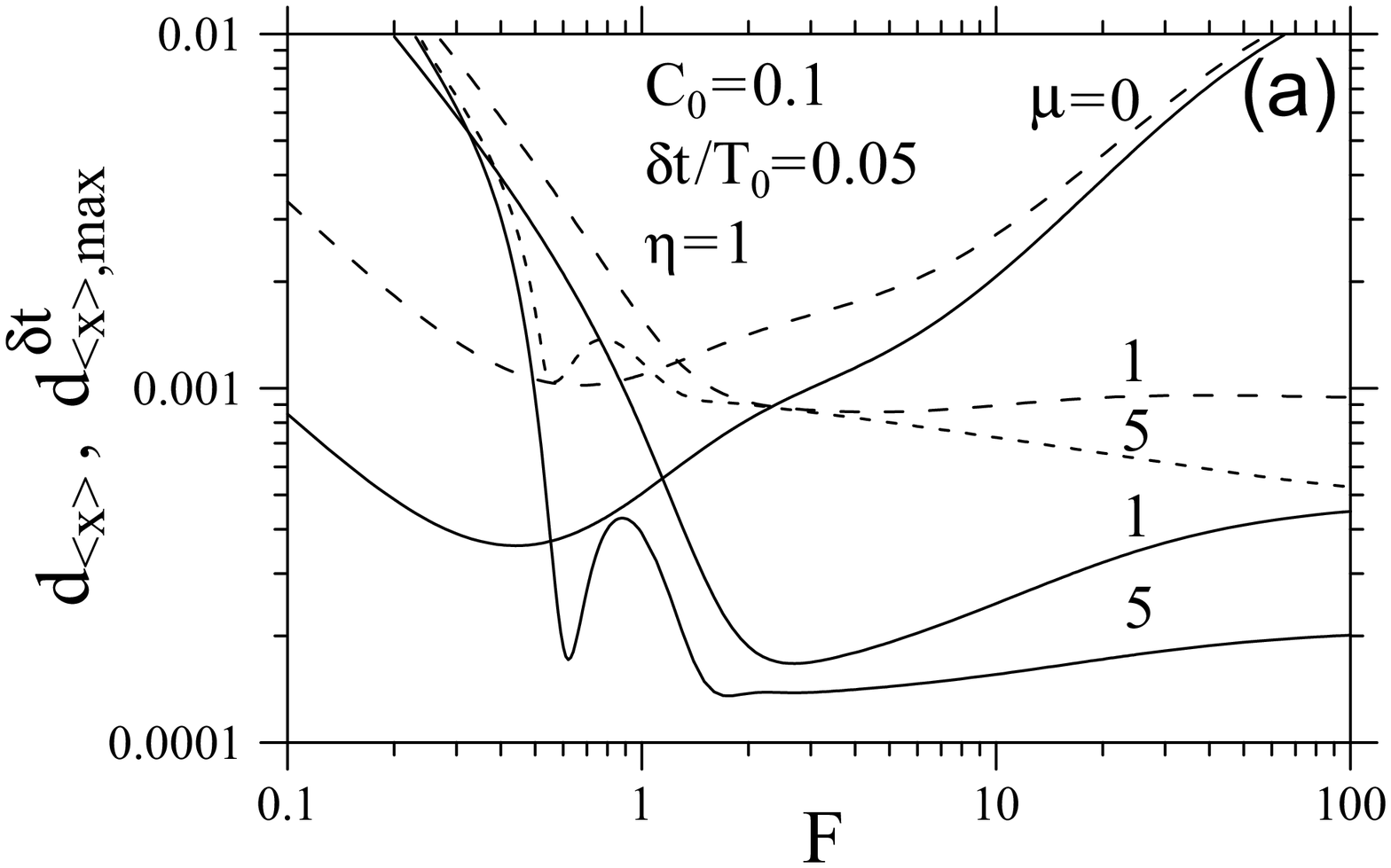}
\includegraphics[width=2.7in]{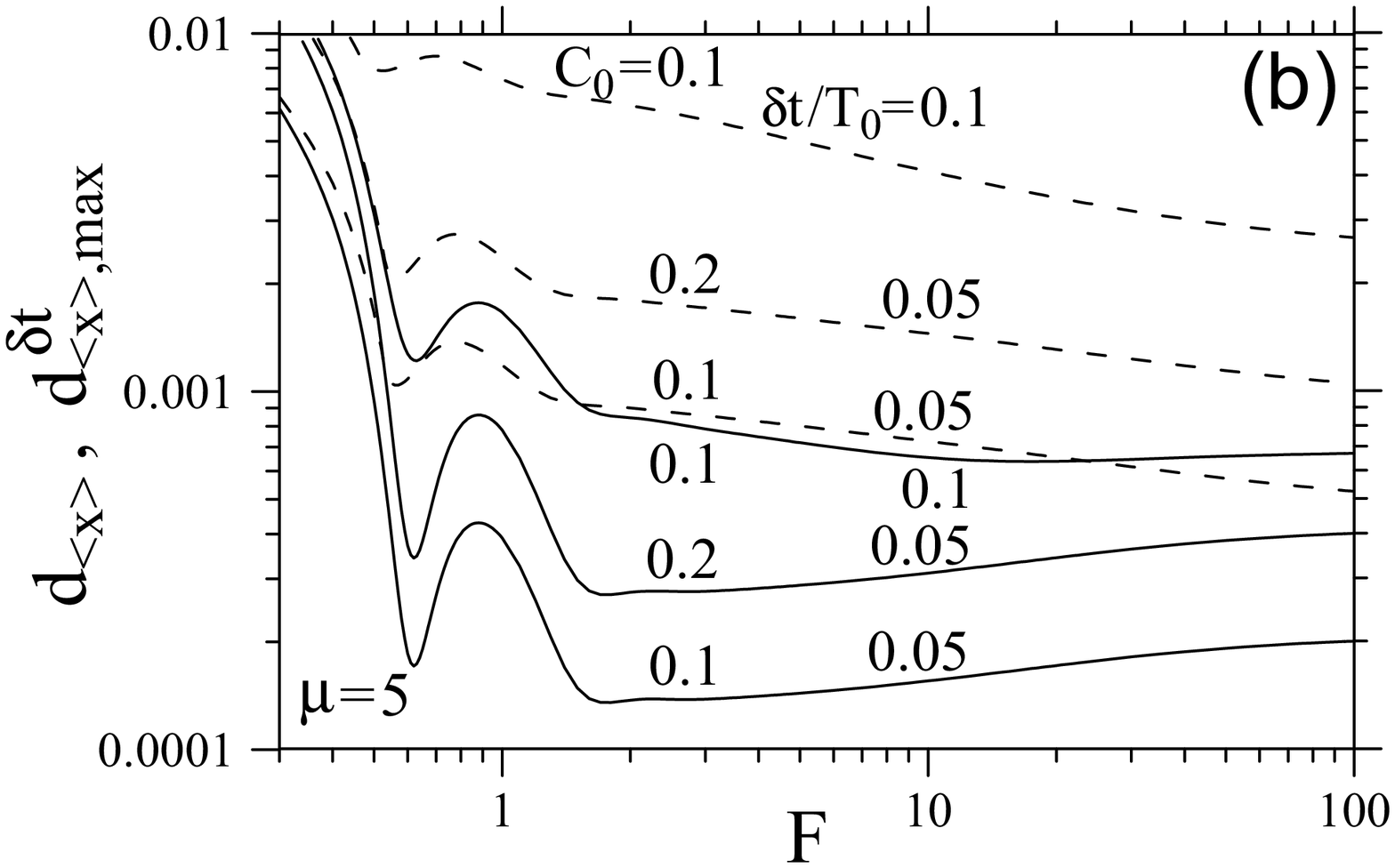}
% where an .eps filename suffix will be assumed under latex,
% and a .pdf suffix will be assumed for pdflatex
\caption{
Variance of the wavepacket center $d_{\langle x\rangle}$ at the
middle of the stroboscopic measurement pulse (solid lines) and
the variance $d_{\langle x\rangle , max}^{\delta t}$ maximized over
the pulse duration (dashed lines), as functions of feedback parameter
$F$ for several values of (a) feedback parameter $\mu$ and (b)
parameters ${\cal C}_0$ and $\delta t$. We assume $Q=\infty$, $\eta =1$,
$\omega = 2\,\omega_0$, and $A_{mod}=1$.
}
\label{fig-fb}
\end{figure}

        We have simulated these equations numerically using the asymptotic
solutions of Eqs.\ (\ref{eq-mDx-dles})--(\ref{eq-mDxp-dles}) for $d_x$,
$d_p$, and $d_{xp}$. We have mostly studied
the resonance $\omega =2\,\omega_0$ in the weakly coupling regime.
Since finite $Q$-factor helps to decrease fluctuations of $\langle x\rangle$,
we have considered only the case $Q=\infty$.
The main finding is that for both harmonic and stroboscopic modulation
of measurement, the center
position variance $d_{\langle x\rangle}$ can be made much smaller than
the packet variance $d_x$ at time moments $t=j2\pi/\omega$ when the packet
squeezing is at maximum.

   Solid lines in Fig. \ref{fig-fb}(a) show stationary values of
$d_{\langle x\rangle}$ at the center of the stroboscopic pulses, as
function of the feedback factor $F$ and several values of feedback factor
$\mu$. The chosen pulse duration $\delta t=0.05\, T_0$ corresponds to
$d_x=1/{\cal S}= 0.091$, while the results for $d_{\langle x\rangle}$
shown in Fig.\ \ref{fig-fb}(a)
are much smaller. One can see that the feedback can operate sufficiently
well for $\mu =0$, so the term with $\gamma_x$ in Eq.\ (\ref{fb-force})
is not really necessary; however, nonzero $\mu$ is beneficial
since it leads to even smaller $d_{\langle x\rangle}$.
The curves in Fig.\ \ref{fig-fb}(a) saturate at $F\rightarrow \infty$,
and the saturation value of $d_{\langle x\rangle}$ decreases
with increase of $\mu$.

        Solid lines in Fig.\ \ref{fig-fb}(b) show the dependence
$d_{\langle x\rangle} (F)$ for a fixed value $\mu =5$ and several
values of the pulse duration $\delta t$ and coupling ${\cal C}_0$.
One can see that $d_{\langle x\rangle}$ decreases with decrease of
both $\delta t$ and ${\cal C}_0$. Since $d_x$ does not depend on ${\cal C}_0$
[see Eq.\ (\ref{S-peak-str})], the ratio $d_{\langle x\rangle} /d_x$
obviously decreases at small coupling.

     The packet center variance $d_{\langle x\rangle} (t)$ changes
significantly within the pulse duration; however, typically it is still
much smaller than $1/{\cal S}$.
Dashed lines in Figs.\ \ref{fig-fb}(a) and \ref{fig-fb}(b) show
$d_{\langle x\rangle}$ maximized over the pulse duration (the maximum
$d_{\langle x\rangle,max}^{\delta t}$ is achieved at the end of pulse).
The dependence of $d_{\langle x\rangle,max}^{\delta t}$ on $F$,
$\mu$, $\delta t$, and ${\cal C}_0$ is generally similar to the
behavior of $d_{\langle x\rangle}$ at the pulse center, though
the values are several times higher.

        For an analytical estimate of $d_{\langle x\rangle}$ let us
assume $F\gg 1$ and $\mu \agt 1$ (we also assume ${\cal C}_0\ll 1$ and
$\delta t/T_0 \ll 1$). Because of the strong damping terms
in Eqs.\ (\ref{eq-cmDx-dles})--(\ref{eq-cmDxp-dles}), the variances
$d_{\langle x\rangle}$, $d_{\langle p\rangle}$, and
$d_{\langle x\rangle \langle p\rangle}$ decay practically to zero
before the start of the measurement pulse.
Within the pulse duration $d_{\langle x\rangle \langle p\rangle}$
can be found from Eq.\ (\ref{eq-cmDxp-dles}) as
$d_{\langle x\rangle \langle p\rangle} \approx -\mu d_{\langle x\rangle }$.
Substituting this value into Eq.\ (\ref{eq-cmDx-dles}) and using
initial condition $d_{\langle x\rangle} (-\delta t /2)=0$ at the
beginning of the pulse, we obtain
$d_{\langle x\rangle}(t) = \omega_0 {\cal C}_0 \int_{-\delta t/2}^{t}
d_x^2(\tau)\,
\exp{\left[- 2 \mu\, \omega_0  \left( t-\tau \right)\right]}\,d\tau$.
Now using the stationary solution $d_x(t) \approx {\cal S}^{-1}
+(\omega_0 t)^2 {\cal S}/\eta$ which follows from
Eq.\ (\ref{dx-str}) for $\varphi =0$ and
${\cal S} \gg 1$,
we can calculate the variance $d_{\langle x \rangle}$ at the pulse center
($t=0$):
\begin{equation}
d_{\langle x\rangle}= \frac{{\cal C}_0 (\omega_0 \delta t)^3}{12\,\eta}
\int_0^{1/2} (1 + 12 y^2)^2
\exp [- 2 \mu y \omega_0 \delta t]\, dy .
        \label{d_xaver-an}
        \end{equation}
In the case $\mu\omega_0\delta t\gg 1$  this expression
simplifies to
        \begin{equation}
d_{\langle x\rangle} ={\cal C}_0 (\omega_0 \delta t)^2/24\mu\eta   ,
        \label{d-fb-1}
        \end{equation}
while in the opposite case $\mu\omega_0\delta t \ll 1$ it gives
        \begin{equation}
d_{\langle x\rangle} ={\cal C}_0 (\omega_0 \delta t)^3/5\eta   .
        \label{d-fb-2}
        \end{equation}
In both cases $d_{\langle x\rangle}$ is much smaller
than ${\cal S}^{-1}=\omega_0 \delta t/2\sqrt{3\eta}$
[see Eq.\ (\ref{S-peak-str})] for
small $\omega_0\delta t$ and/or small ${\cal C}_0/\sqrt{\eta}$.

    It is easy to see that the maximum value of $d_{\langle x\rangle}(t)$
within the pulse duration is achieved at its end ($t=\delta t/2$),
and can be calculated by Eq.\ (\ref{d_xaver-an}) with the lower
integration limit extended to $y=-1/2$ and with extra factor
$\exp (-\mu\omega_0\delta t)$.
In particular, this gives
$d_{\langle x\rangle , max}^{\delta t}=16\,d_{\langle x\rangle}$ for
$\mu\omega_0\delta t\gg 1$
and $d_{\langle x\rangle , max}^{\delta t} = 2\,d_{\langle x\rangle}$
for $\mu\omega_0\delta t\ll 1$,
which confirms numerical result and shows that
$d_{\langle x\rangle , max}^{\delta t}$ can also be made much smaller than
$d_x=1/{\cal S}$, similar to the result for $d_{\langle x\rangle}$.

        Overall, the analytical and numerical results show that
the feedback is sufficiently efficient for a broad range of feedback
parameters $F$ and $\mu$.

        At the end of this Section we would like to discuss the
following concern on possibility to use quantum feedback in case
of stroboscopic measurements. The general idea of stroboscopic
measurement is to avoid obtaining any information on phase of the
nanoresonator oscillations, while quantum feedback requires to
know the phase of packet center oscillations. So, a natural
question is how it happens that we monitor this phase.

        A qualitative answer is that once we know $\langle x\rangle$
and $\langle p\rangle$, their further evolution can be extracted
from the measurement record $I(t)$ via Eqs.\ (\ref{eq-m-x})--(\ref{eq-m-p})
even though the measurement is performed during only small fraction of the
period. Initial knowledge of $\langle x\rangle$ and $\langle p\rangle$
can be eventually obtained also using Eqs.\ (\ref{eq-m-x})--(\ref{eq-m-p})
starting from any (incorrect) initial condition, since the solution
of the equations gradually forgets initial condition and is eventually
dominated by the known noise term.

Let us assume that we start using
Eqs.\ (\ref{eq-m-x})--(\ref{eq-m-p}) with incorrect initial conditions
$\langle x_1(0)\rangle$ and $\langle p_1(0)\rangle$ instead of correct
values $\langle x_2(0)\rangle$ and $\langle p_2(0)\rangle$, and let us
show that the normalized differences between the corresponding solutions
$\tilde x=(\langle x_1\rangle- \langle x_2\rangle )/\sqrt{\hbar /2m\omega_0}$
and
$\tilde p=(\langle p_1\rangle- \langle p_2\rangle )/\sqrt{\hbar m\omega_0/2}$
decay to zero with time.
Since the same measurement record $I(t)$ is used for both
solutions, the  value of the noise term $\xi =I-\langle I \rangle$
is affected by $\tilde x$, and the difference evolves as
        \begin{eqnarray}
&& d {\tilde x}/dt = \omega_0 [ \tilde p-{\cal C}_0 |f(t)| d_x \tilde x ] ,
        \\
&& d{\tilde p}/dt = \omega_0 [ -\tilde x-{\cal C}_0 |f(t)| d_{xp}\, \tilde x
        -\tilde p /Q].
        \end{eqnarray}
Finite $Q$-factor obviously damps oscillations of $\tilde x$ and $\tilde p$,
so for a worst case let us assume $Q=\infty$.
Then the evolution of the ``energy function''
$\tilde \varepsilon = {\tilde x}^2+{\tilde p}^2$
is
$d{\tilde \varepsilon}/dt = -2\omega_0 {\cal C}_0 |f(t)| (d_x {\tilde x}^2
   + d_{xp} {\tilde x}{\tilde p})$.
Assuming weak coupling,
 using asymptotic time dependence of variances
$d_x={\cal S}^{-1} +A[1-\cos (2\omega_0 t)]$,
$d_{xp} = A\sin (2\omega_0 t)$ [see Eqs.\ (\ref{dx-str}) and
(\ref{dxp-str})],
and assuming $\tilde x=\tilde \varepsilon^{1/2} \sin (\omega_0 t+\phi )$,
$\tilde p=\tilde \varepsilon^{1/2} \cos (\omega_0 t+\phi )$,
we derive equation
$d\tilde \varepsilon /dt = -\tilde\varepsilon \omega_0 {\cal C}_0 |f(t)|
\{ (A+ {\cal S}^{-1}) [1-\cos (2\omega_0 t+2\phi )]
-A[ \cos (2\omega_0 t) -\cos (2\phi )] \}$.
After averaging over the short pulse duration $\delta t$, the expression
in curly brackets becomes
$A[1+\cos (2\phi )](\omega_0\delta t)^2/6 +
{\cal S}^{-1}[1-\cos (2\phi)(1-(\omega_0\delta t)^2/6)]$, which
is always positive.
Therefore, $\tilde\varepsilon$ decays to zero, and this
happens on the timescale
$\sim T_0 \eta^{1/2}{\cal C}_0^{-1}(\omega_0 \delta t)^{-2}$,
comparable to the timescale of purity saturation and squeezing
buildup (see Section IV.C).
So, we have proven that $\langle x\rangle$ and
$\langle p\rangle$ calculated from Eqs.\ (\ref{eq-m-x})--(\ref{eq-m-p})
eventually
depend only on the measurement record and do not depend on initial values.
As a by-product, this statement also means that a mixture of Gaussian
states (which in general is not Gaussian) eventually becomes a single
Gaussian state.

\section{Verification of  squeezed state}

        The fact that the squeezed state of a nanoresonator can be prepared
by the modulated measurement and quantum feedback, does not automatically
mean that this state may be useful for the measurement of extremely weak
forces, and even that such state can be checked experimentally in a
straightforward way. As an example of such problem, in one of setups analyzed
in Ref.\ \onlinecite{Wiseman-94} the squeezed in-loop optical state is
realized by using quantum feedback, but the squeezing of the output light is
impossible. Fortunately, as we discuss below, in our case there is no problem
with observability of the squeezed state.

We have studied the possibility to verify the squeezed state of the
nanoresonator in the following way. After the preparation of the squeezed
state by stroboscopic measurement and feedback, the feedback
at some moment ($t=0$) is switched off, while the stroboscopic measurement
continues. Considering for simplicity the case of one measurement per
nanoresonator period ($n=2$, $\omega =\omega_0$), we  average
the position measurement result $x_j^m$ for $j$th pulse over many pulses
 (each pulse gives a very imprecise result
because of weak coupling):
        \begin{equation}
X_N = \frac{1}{N} \sum_{j=1}^N x_j^m
= \frac{1}{N} \sum_{j=1}^N \frac{1}{\delta t \, k_0}
 \int_{jT_0-\delta t/2}^{jT_0+\delta t/2} [I(t)-I_{00}] \, dt \, .
        \label{X_N}
        \end{equation}
As we show below, for a squeezed initial state, the r.m.s.\ fluctuation
of $X_N$ can be much smaller than if we would start with the ground state
(and much smaller than $\Delta x_0$).
This is the way to verify squeezing, and also this procedure is exactly
what can be used for an ultrasensitive force measurement with accuracy
beyond the standard quantum limit.
[Notice that
for two measurements per period, $\omega =2\,\omega_0$, the definition
(\ref{X_N})
should be modified by adding odd (``$\pi$-phase'') contributions
with negative sign. Then all results of this Section are valid for
$\omega =2\,\omega_0$ as well.] For simplicity in this Section we
neglect the effect of finite quality factor $Q$ of the nanoresonator.

        The analysis of the distribution of $X_N$ (over realizations)
is very simple in the case of instantaneous but imprecise measurements,
$\delta t\rightarrow 0$, ${\cal C}_0\delta t = \mbox{const}$, since the
Hamiltonian evolution of the resonator in between the measurements can be
completely neglected. Therefore the problem reduces to a classical
sequential measurement of a ``particle'' position, which is initially
characterized by the Gaussian probability distribution with variance
$\Delta x_0^2/{\cal S}$ (recall $\Delta x_0=\sqrt{\hbar /2m\omega_0}$),
while each imprecise measurement has
variance $(\Delta x_0)^2/ {\cal C}_0 \omega_0 \delta t$.
In particular, $N$ measurements with results $x_j^m$ are equivalent to one
$N$-times stronger measurement with result $X_N$ [mathematically this is
because the product of several
measurement Gaussians as in Eq.\ (\ref{Bayes-diag}) is the Gaussian with
added inverse variances and centered at $X_N$].
Then distribution of $X_N$ is the convolution of the initial state Gaussian
and the total measurement Gaussian [see Eq.\ (\ref{P(I)})];
so the variance of $X_N$ is equal to the sum of corresponding variances:
        \begin{equation}
D_{X,N} = \Delta x_0^2 \left( \frac{1}{\cal S} +
        \frac{1}{N{\cal C}_0 \omega_0 \delta t} \right)  .
        \label{DXN}
        \end{equation}
For completeness let us also mention that after $N$ measurements the
``actual'' position is characterized by the Gaussian probability
[see Eq.\ (\ref{Bayes-diag})] with variance
$\Delta x_0^2 ({\cal S}+ N{\cal C}_0\omega_0\delta t)^{-1}$
(inverse variances are added for product of Gaussians) and
centered at $X_N /(1+{\cal S}/N{\cal C}_0\omega_0\delta t)$,
which is the weighted sum of the initial center of distribution (assumed
to be zero) and the measurement result $X_N$.

        Obviously, at $N \gg 1/{\cal C}_0\omega_0\delta t$ the variance
of $X_N$ given by Eq.\ (\ref{DXN}) is significantly smaller for a squeezed
state (${\cal S}>1$) than for the
ground state (${\cal S}=1$). Even though this difference can be
rigorously verified only by performing many experiments to accumulate
statistics for $D_{X,N}$, it can be observed even in a single experiment
with good reliability if ${\cal S}\gg 1$ (for applications like
force detection we should discuss single realizations).
The error probability for distinguishing the two cases in one
trial is essentially the overlap of two distributions for $X_N$, which is
crudely ${\cal S}^{-1/2}$ for $N\rightarrow \infty$
(ratio of distribution widths).
[A better approximation for error probability to distinguish two Gaussians
with coinciding centers and different variances $D_1$ and $D_2$
is $(\mbox{ln}R / 2\pi R)^{1/2}$ where $R=D_1/D_2 \gg 1$;
in our case $R={\cal S}$.]
So, the squeezed state with ${\cal S}\gg 1$ can be reliably verified
even in a single experiment.

        Unfortunately, this result requires the assumption of infinitely
strong coupling with detector (${\cal C}_0\rightarrow \infty$),
so it is not obvious if it holds in the practical case of weak coupling
(${\cal C}_0\ll 1$)
or not. The anticipated problem is that for sufficiently large $N$
which makes the second term in Eq.\ (\ref{DXN}) sufficiently small,
the nanoresonator
heating due to measurement back-action may already eliminate the squeezing
(the feedback is off).
To resolve this issue we have calculated $D_{X,N}$ for stroboscopic
modulation numerically by Monte-Carlo simulation of realizations using
Eqs.\ (\ref{eq-m-x})--(\ref{eq-m-p}) and then averaging over realizations.
Such simulation happened to be not too simple; in particular, the
time step should be chosen carefully. As a check of simulation accuracy
we were comparing the variance of $\langle x\rangle$ obtained by
averaging over Monte-Carlo realizations with the results from Eqs.\
(\ref{eq-cmDx-dles})--(\ref{eq-cmDxp-dles}) without feedback; the
difference was checked to be within few percent.

\begin{figure}
\centering
\includegraphics[width=2.7in]{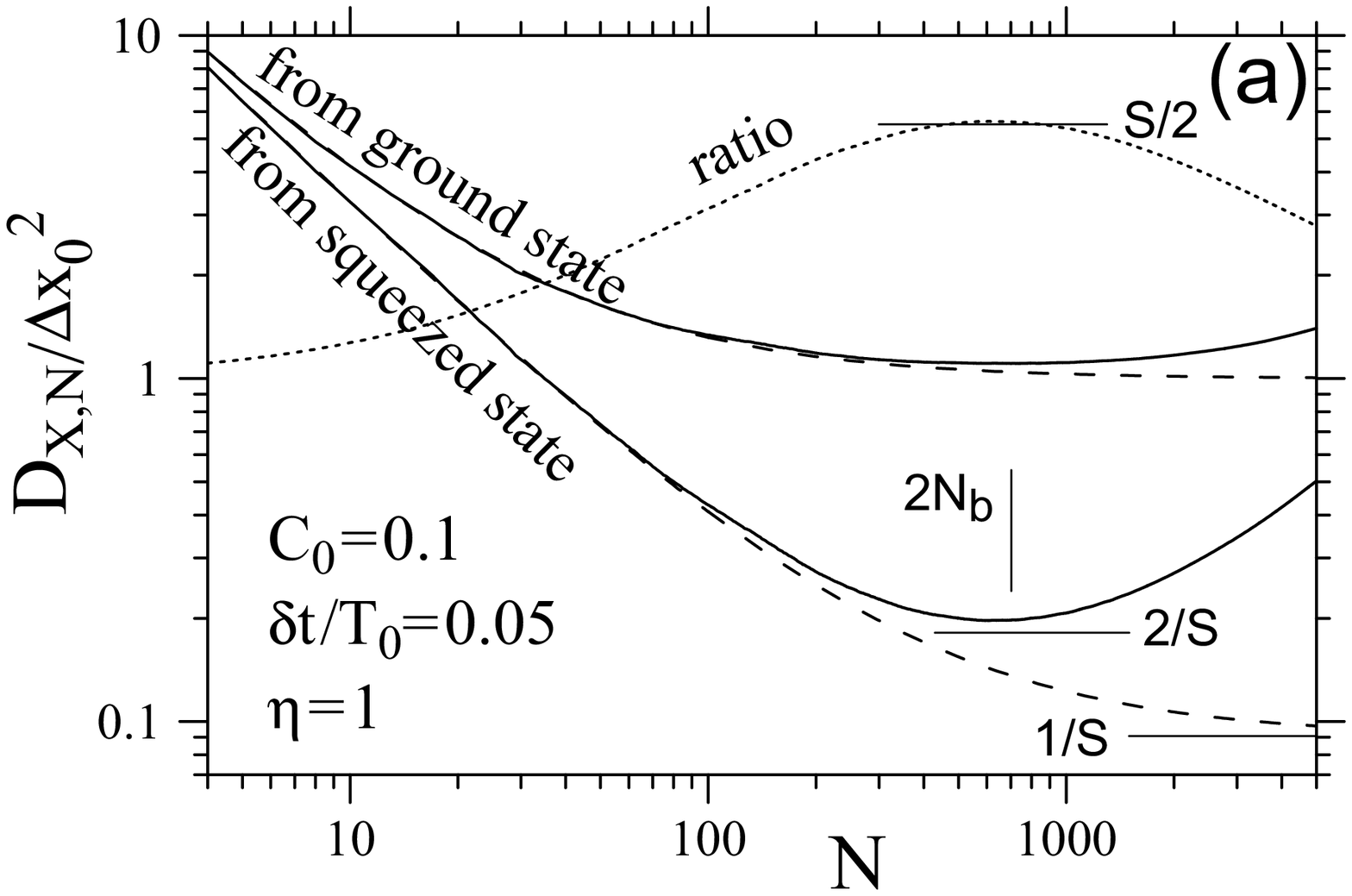}
\includegraphics[width=2.7in]{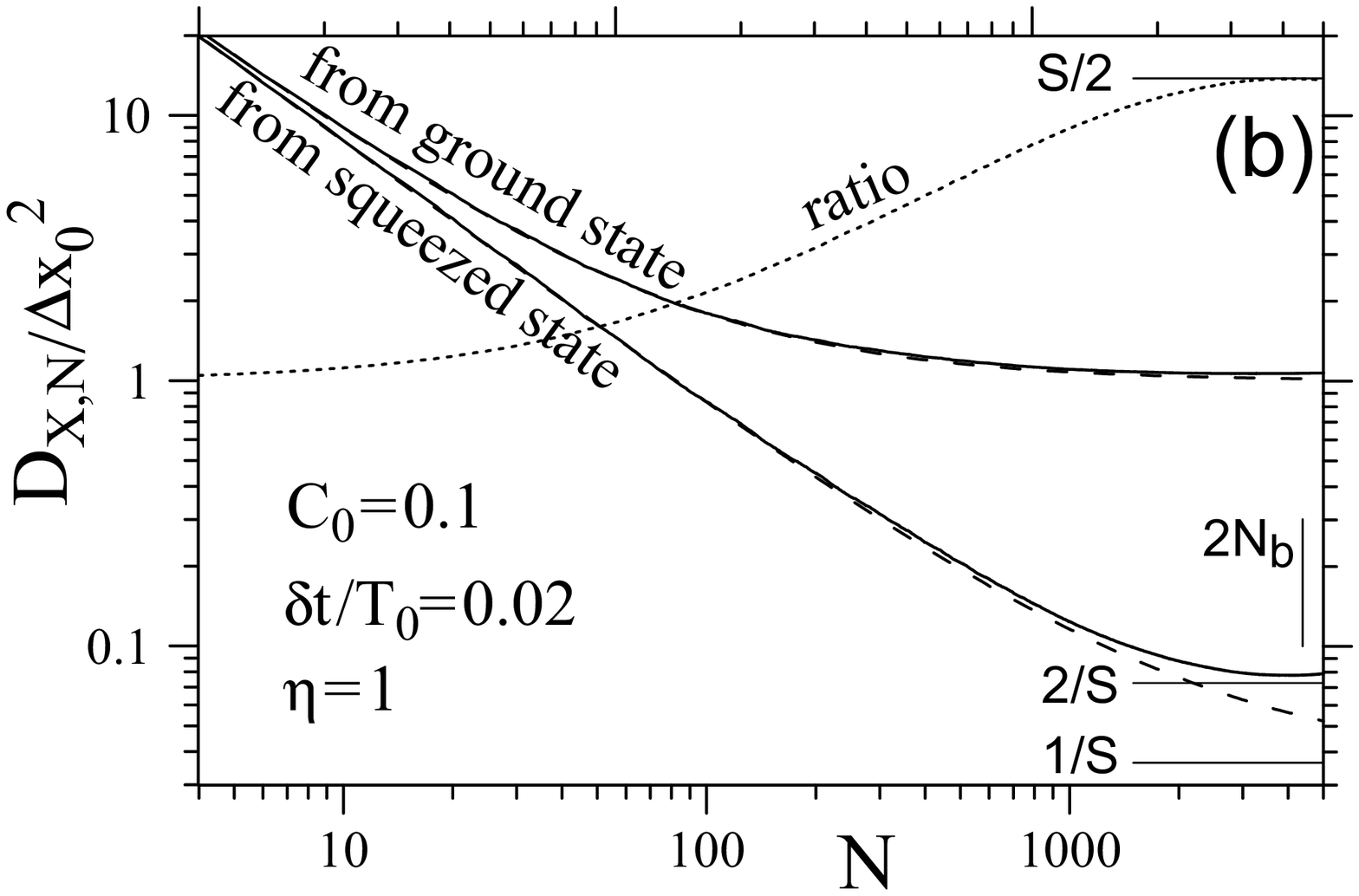}
% where an .eps filename suffix will be assumed under latex,
% and a .pdf suffix will be assumed for pdflatex
\caption{
Variance $D_{X,N}$ of the measurement result $X_N$ [see Eq.\ (\ref{X_N})]
as a function of number $N$ of stroboscopic measurement pulses.
The measurement
procedure is applied either to the ground state or to the squeezed state
prepared by the same procedure complemented with quantum feedback.
Panels (a) and (b) are for different durations of the measurement pulses,
corresponding to initial squeezing ${\cal S}=11.0$ and ${\cal S}=27.6$.
Solid lines are the numerical results for finite pulse duration $\delta t$,
while
dashed lines correspond to Eq.\ (\ref{DXN}) (instantaneous measurements).
Dotted lines are the ratios of the results shown by solid lines.
Force detection beyond the standard quantum limit is possible when
$D_{X,N}/\Delta x_0^2 <1$.
}
\label{fig-obs}
\end{figure}

        Solid lines in Fig.\ \ref{fig-obs} show the numerical results
for $D_{X,N}$ for weak coupling (${\cal C}_0=0.1$) and two values of
pulse duration: (a) $\delta t/T_0=0.05$ and (b) $\delta t/T_0=0.02$.
Initial state is either ground state or asymptotic zero-centered squeezed
state corresponding to the same measurement parameters, so that the squeezing
is given by Eq.\ (\ref{S-peak-str}) and preparation of the squeezed state
differs from its verification only by quantum feedback switched on or off.
 Dashed lines in Fig.\ \ref{fig-obs} are calculated
using Eq.\ (\ref{DXN}). One can see that the numerical results follow
the simple analytics when the contribution from the measurement accuracy
in Eq.\ (\ref{DXN}) dominates; however, at larger number of pulses $N$
the numerical results deviate upwards and eventually $D_{X,N}$ starts
to increase with $N$, which is expected because of the nanoresonator
``heating'' due to measurement back-action.

The numerical minimum of $D_{X,N}$ for squeezed states (${\cal S}\gg 1$)
in Fig.\ \ref{fig-obs} is a little higher than $(2/{\cal S})\Delta x_0^2$.
We have checked that the minimum is still close to $(2/{\cal S})\Delta x_0^2$
for several other values
of ${\cal C}_0$ and $\delta t$. As seen in Fig.\ \ref{fig-obs}, this
minimum is achieved at $N$ close to $2N_b$, where $N_b$ given by
Eq.\ (\ref{N_b}) is the estimate of number of measurement pulses
for squeezing buildup. We have checked that this result also
holds for different values of ${\cal C}_0$ and ${\delta t}$.
The fact that the minimum of $D_{X,N}$ is higher than
$(2/{\cal S})\Delta x_0^2$
is not surprising since the average squared wavepacket width $D_x$ within the
pulse duration $\delta t$ is $(2/{\cal S})\Delta x_0^2$ for ${\cal S}\gg 1$
(see Section IV.B.3).
Hence, one could even guess that $D_{X,N}$ should be always larger
than Eq.\ (\ref{DXN}) with $1/{\cal S}$ replaced by $2/{\cal S}$.
However, actually $D_{X,N}$ goes below such a bound for a range of $N$.
Some understanding of this fact can be provided by an argument that
for a classical measurement the nanoresonator motion
during $\delta t$ would be averaged and so $X_{N}$ would depend only
on the positions at the centers of the measurement pulses.

The minimum of $D_{X,N}$ for the case when we start measurement procedure
from the ground state, is only a little larger than $\Delta x_0^2$
(see Fig.\ \ref{fig-obs}),
which means that  the accumulated measurement accuracy becomes
better than the standard quantum limit $\Delta x_0$ at sufficiently
smaller $N$ than when the back-action heating becomes important.
 Therefore, the ratio of
$D_{X,N}$ starting with the ground and squeezed states (dotted lines in
Fig.\ \ref{fig-obs}) reaches the maximum of approximately ${\cal S}/2$
[in the case of instantaneous measurements described by Eq.\ (\ref{DXN}),
this ratio would approach $\cal S$ at $N\rightarrow \infty$].

        Thus, our numerical results show that for a proper duration
of the measurement procedure ($\sim N_bT_0$) the variances $D_{X,N}$
for the squeezed and ground initial states are significantly different,
and therefore these states can be reliably distinguished.
The squeezed state verification using a weakly coupled detector is
only by a factor $\sim 2$ less efficient than similar procedure using
instantaneous measurements by strongly coupled detector.
Even though these results have been obtained neglecting the effect of
the resonator $Q$-factor, we do not expect a significant difference
for finite $Q$ because it affects equally the preparation of the squeezed
state and its verification.
Finally, let us mention that if an external force has shifted the
nanoresonator position by $\Delta x$, the procedure discussed in this
Section can detect the force if $\Delta x \agt \sqrt{2/{\cal S}}\Delta x_0$.

\section{Conclusion}

        As analyzed in this paper, the uncertainty of the
nanoresonator position can be squeezed significantly below the ground state
level by using the modulated in time ($\omega \approx 2\,\omega_0/n$)
continuous measurement of the nanoresonator
position with the QPC or RF-SET detector. The measurement strength can be
modulated by applying the periodic voltage across the detector.
For the RF-SET the modulation can also be done by varying the gate voltage;
however, it is important that such modulation periodically brings the SET
in the Coulomb blockade regime, so the back-action is periodically switched
off (actually, similar gate voltage modulation is also possible for the QPC,
but it is quite unnatural). The mechanism of squeezing is similar to the
stroboscopic QND measurements: \cite{BraginskyKhalili,Braginsky2,Thorne}
for periodic measurement pulses separated by integer number of
half-periods of oscillation, the free evolution of the resonator is
to the large extent compensated, which allows the buildup of the
effective measurement strength for repeated imprecise measurements;
therefore the squeezed state is produced when effective measurement
accuracy becomes better than the ground state width $\Delta x_0$.
A significant difference between our analysis and the standard QND case
of instantaneous stroboscopic measurements is the assumption of weak
coupling with detector, ${\cal C}_0\lesssim 1$, while ${\cal C}_0$
should be infinitely large for instantaneous measurements.
Obviously, the squeezed state oscillates with time, so that the moments of
minimum position uncertainty $\Delta x_0/\sqrt{\cal S}$ and the
minimum momentum uncertainty $\hbar /2\Delta x_0 \sqrt{\cal S}$
are shifted in time by $T_0/4=\pi/2\omega_0$.

        We have considered harmonic (\ref{harm-mod}) and stroboscopic
(\ref{strob-mod}) modulations with frequency $\omega$ and modulation
amplitude $A_{mod}$. As anticipated, $A_{mod}=1$ is found to be
the optimum value for maximum squeezing in both cases.
We have found that only a moderate squeezing ${\cal S} = \sqrt{3\eta}$
(requiring relatively high detector quantum efficiency $\eta$)
is possible for the harmonic modulation with
twice the resonator frequency, $\omega = 2\, \omega_0$
[see Eqs.\ (\ref{S-max-harm}), (\ref{S-A}), (\ref{Sq-A-cos}),
and Fig.\ \ref{fig-cos}]. In contrast,
an arbitrary strong squeezing is in principle possible for the
stroboscopic modulation when the measurement (and therefore back-action)
is switched completely off in between measurement pulses of short duration
$\delta t$. If not limited by effects of resonator quality factor $Q$,
the squeezing can be up to
${\cal S} = 2\sqrt{3\eta}/\omega_0\delta t$
at frequency $\omega =2\,\omega_0/n$ [see Eqs.\
(\ref{Sq-max-strob})--(\ref{S-width-str}) and Figs.\ \ref{fig-strob1} and
\ref{fig-strob2}].
The squeezing buildup requires on the order of
$\sqrt{\eta}/{\cal C}_0 (\omega_0\delta t)^2$ measurement pulses
[see Eq.\ (\ref{N_b})], so for a limited ``waiting time'' $\tau_w$
the squeezing cannot exceed ${\cal S} \simeq 2 \eta^{1/4}
({\cal C}_0\tau_w/T_0)^{1/2}$ [see Fig.\ \ref{tau_w} and Eqs.\
(\ref{S-tau_w1})--(\ref{S-tau_w2})].
       Finite $Q$-factor of the nanoresonator limits the squeezing
by ${\cal S}=[(2/n)(\omega_0\delta t/2\pi)\, {\cal C}_0 Q /
\mbox{coth}(\hbar\omega_0 /2T)]^{1/2}$ [see Eq.\ (\ref{S-Q-an})];
after optimization over $\delta t$ this leads to the limitation
 ${\cal S}\simeq (3/4)\eta^{1/6}[{\cal C}_0 Q/n\,\mbox{coth}
(\hbar \omega_0/2T)]^{1/3}$ [see Fig.\ \ref{fig-Q} and Eq.\ (\ref{Q-fit})].
Notice that this result is consistent with the mentioned in the Introduction
condition of quantum behavior \cite{BraginskyKhalili}
 $T\tau_m/Q\lesssim \hbar$ for a good detector,
$\eta \sim 1$, and measurement time $\tau_m=4/{\cal C}_0\omega_0$
corresponding to $x$-accuracy equal to $\Delta x_0$.

        While the modulated measurement squeezes the width of the
resonator wavepacket, the position of its center $\langle x\rangle$
fluctuates due to random back-action from the detector, and may
deviate very far away from the origin. To keep the packet center near
$x=0$ we apply quantum feedback similar to Refs.\
\onlinecite{DohertyJacobs} and \onlinecite{Hopkins}
(the packet center in momentum space in this case will be kept near
zero as well). We have found [see Fig.\ \ref{fig-fb} and
Eqs.\ (\ref{d-fb-1}) and (\ref{d-fb-2})] that the feedback
can keep the deviation of $\langle x\rangle$ from zero much smaller
than the packet width $\Delta x_0/\sqrt{\cal S}$,
which means that the ensemble-averaged squeezing
practically does not differ from the packet width squeezing.

     Verification of the squeezed state can be performed in essentially
the same way as its preparation, the only difference is that the quantum
feedback should be switched off. We have studied the distribution of
the position measurement result $X_{N}$ averaged over $N$ stroboscopic
measurement pulses and found (see Fig.\ \ref{fig-obs})
that for a significant range of $N$ before the back-action heating
becomes important, the width of $X_N$ distribution is close
to $\sqrt{2/{\cal S}}\Delta x_0$, which may be much smaller than the
ground state width $\Delta x_0$.
The analyzed procedure can be applied in a straightforward way for
ultrasensitive force detection beyond the standard quantum limit:
the force can be detected when it causes the nanoresonator shift
$\Delta x$ larger than $\sqrt{2/{\cal S}}\Delta x_0$.

        For an estimate of the present-day experimental parameters
let us use the data from Ref.\ \onlinecite{Schwab-QL}.
The experimental sensitivity of $3.8 \, \mbox{fm}/ \sqrt{\mbox{Hz}}$
for the nanoresonator with $\omega_0/2\pi =19.7$ MHz and $\Delta x_0 =21$ fm
can be translated into the dimensionless coupling ${\cal C}_0 \simeq
5\times 10^{-7}$. For the $Q$-factor of $3.5\times 10^4$ and using a crude
estimate for quantum efficiency $\eta \sim 10^{-1}$, we have the product
${\cal C}_0 Q\sqrt{\eta} \simeq 6\times 10^{-2}$. Since this product
should be larger than at least 10 for a noticeable squeezing (see Fig.\
\ref{fig-Q}), we should conclude that it is still 2-3 orders of magnitude
less than needed for squeezing.
However, the necessary improvement of experimental parameters may be
reachable in reasonably near future (notice that ${\cal C}_0$ scales
quadratically with response $k_0$; the estimates of Ref.\
\onlinecite{Hopkins} give ${\cal C}_0\simeq 10^{-3}$).
 For a reasonably realistic
parameters ${\cal C}_0 \sim 10^{-2}$, $Q\sim 10^{6}$, and $\eta \sim 0.3$
the product ${\cal C}_0 Q\sqrt{\eta} \simeq 5\times 10^3$, therefore
the low-temperature squeezing  ${\cal S}\simeq 13$ is possible,
and a significant squeezing survives up to temperatures
$T\sim 10 \,\hbar\omega_0$.

    In an experiment it may be convenient to flip every second time the sign
of stroboscopic voltage pulse applied to the detector. Then the information
on average position $X_N$ [see Eq.\ (\ref{X_N})] can be extracted
from the low-frequency component of the detector current
(somewhat similar to the RF-SET mixer of Ref.\ \onlinecite{Cleland-2003}).
Even though we expect that the high-frequency component would still
be necessary
for quantum feedback, the results of Section VI indicate that the
preparation-detection procedure should work reasonably well even without
feedback if the preparation time is comparable to the squeezing
buildup time (so that the back-action heating is not yet too strong).

        Concluding, we hope that the QND squeezing of a nanoresonator
can be demonstrated experimentally in a reasonably near future and will
eventually be useful for the force detection with sensitivity beyond
the standard quantum limit.

%        \section*{Acknowledgment}

The authors would like to thank D. Averin, A. Doherty, S. Habib,
K. Jacobs, K. Likharev, I. Martin, and G. Milburn
for fruitful discussions and remarks.
The work was supported by NSA and ARDA under ARO grants DAAD19-01-1-0491
and W911NF-04-1-0204  (R.R.\ and A.K.)  and by NSA (K.S.).

\begin{appendix}*
\section{Generalized Bayesian formalism for a nanoresonator}

In this Appendix we generalize the Bayesian equation (\ref{meas-Ito-linear})
to the case of a detector with correlation between output and back-action
noises, and also discuss the contributions from various kinds of the noise.
We discuss only the nanoresonator evolution due to
measurement; therefore the terms ${\cal H}_0$, ${\cal H}_{env}$, and
${\cal H}_{fb}$ in the Hamiltonian (\ref{Hamiltonian}) are neglected.
For simplicity we also do not consider the modulation of measurement
parameters.

        Following the logic of Ref.\ \onlinecite{Kor-nonid}, we
discuss first the effects of several additional classical noises.
Let us start with additional classical white noise $\xi_1(t)$ at the output,
so that the total output noise $\xi = \xi_{id} +\xi_1$ consists
of the ``ideal quantum contribution'' $\xi_{id}$ discussed in Section II
[see Eq.\ (\ref{dif-eq3})] and $\xi_1$; the corresponding spectral
densities are $S_I=S_{id}+S_1$. Using the ``double Bayesian''
procedure of Ref.\ \onlinecite{Kor-nonid} it is simple to show that
averaging over $\xi_1$ leads to the addition of the decoherence term
$-\gamma_1 (x-x')^2 \rho(x,x')$ with $\gamma_1=k^2S_1/4S_{id}S_I$
into Eq.\ (\ref{dif-eq3}). Therefore, the effect of $\xi_1$
is the reduction of the quantum efficiency $\eta$ from ideal value
$\eta =1$ to $\eta =S_{id}/S_I$.

        The second natural noise source is the classical force
$\xi_2(t)$ (uncorrelated with $\xi_1$) with white spectral density $S_2$,
which leads to
the stochastic term $-\xi_2 (t){\hat x}$ in the Hamiltonian.
Averaging over $\xi_2$ gives the extra decoherence term
$-\gamma_2 (x-x')^2 \rho(x,x')$ in Eq.\ (\ref{dif-eq3})
with $\gamma_2=S_2/4\hbar^2$. Therefore the effect of
$\xi_2$ can still be taken into account by further reduction
of the efficiency $\eta$.

       When the nanoresonator is measured by a single-electron transistor,
the back-action force is in general correlated with the output noise.
To take this correlation into account, let us introduce one more
stochastic classical force $\xi_3(t)=\alpha \xi_1(t)$ fully correlated
with output noise $\xi_1$ (this obviously accounts for arbitrary
correlation between the total force $\xi_2+\xi_3$ and $\xi_1$).
Averaging over $\xi_3$ leads to the terms
        \begin{equation}
i{\cal K} (x-x') \rho(x,x') \xi (t) - (\gamma_3 +{\cal K}^2 S_I/4)
 (x-x')^2 \rho(x,x')
        \label{correl-terms}
        \end{equation}
with correlation factor ${\cal K}=\alpha S_1/\hbar S_I$ and
decoherence  $\gamma_3=\alpha^2 S_{id}S_1/4\hbar^2S_I$
to be added into Eq.\ (\ref{dif-eq3}).
[Notice that  Eq.\ (\ref{dif-eq3}) is in the It\^o form; there is no
contribution ${\cal K}^2 S_I/4$ to decoherence in the Stratonovich form.]
The correlation term cannot be described in terms of efficiency $\eta$
and requires generalization of the Bayesian equation (\ref{meas-Ito-linear}).

        For measurement by single-electron transistor the average back-action
force actually depends on the nanoresonator position $x$ in a rather
complicated way, and this leads to additional potential energy term
${\cal V}_{add}({\hat x})$ in the Hamiltonian. In general this term
contributes to unharmonicity of the nanoresonator, though for small
amplitude of oscillations it mainly shifts the equilibrium point and
renormalizes the spring constant.

        The effects of correlation between the output and back-action
noises are also important for a detector with ``asymmetric'' coupling
described by nonzero relative phase \cite{Kor-Av,GoanMilburn,Kor-nonid}
between complex magnitudes $M$ and $\Delta M$ in the Hamiltonian terms
(\ref{Hdet}) and (\ref{Hint}). Evolution equation (\ref{dif-eq3})
for such a detector should be complemented \cite{GoanMilburn,Kor-nonid}
by the terms similar to Eq.\ (\ref{correl-terms}) with correlation
factor ${\cal K} =e^{-1} |\Delta M/M| \sin [\arg (\Delta M/M)]$ but
without dephasing, $\gamma_3=0$ (the detector is still ideal in the sense
that a pure state of the nanoresonator remains pure in the course
of measurement).
   Besides the correlation term, the oscillator potential is changed
by the contribution ${\cal V}_{add}(x) = -\hbar {\cal K}(I_0+kx)^2/2k
+\mbox{const}$.

        For completeness let us also consider the noise of the spring
constant described by the stochastic potential energy $\xi_4 x^2$.
 Averaging over this noise leads to the term
$-\gamma_{spr} (x^2-x'^2)^2 \rho (x,x')$
(where $\gamma_{spr}=S_4/4\hbar^2$)  which has significantly different
form compared to the standard decoherence term (in particular, this
term makes the density matrix non-Gaussian).

        Combining all contributions,
the nanoresonator evolution due to measurement is described
in It\^o form  as

        \begin{eqnarray}
&& \hspace{-.5cm}
\dot{\rho}(x,x') =
 -\left( \frac{k^2}{4S_I} + \frac{{\cal K}^2 S_I}{4} +\gamma_d
\right) (x-x')^2 \, \rho(x,x')
        \nonumber\\
        && \hspace{-0.1cm}
+\left(
\frac{k}{S_I} \, (x+x'-2\langle x\rangle )
+ i {\cal K} (x-x') \right)
\rho(x,x') \, \xi(t)
        \nonumber\\
        && \hspace{-0.1cm}
+ [{\cal V}_{add}({\hat x}),\rho ]_{x,x'}
-\gamma_{spr} (x^2-x'^2)^2 \rho (x,x') \, ,
        \label{meas-gener}
        \end{eqnarray}
where ${\cal K}$ is the total correlation factor, $\gamma_d$ is the total
dephasing, $\gamma_{spr}$ is due to noise of the spring constant,
and ${\cal V}_{add}(x)$ is the renormalization of the resonator
potential energy.

\end{appendix}

\end{document}